\documentclass{jfm}

\usepackage{graphicx}
\usepackage{epstopdf, epsfig}
\usepackage{psfrag}
\usepackage{bm}
\usepackage{color}

\newcommand{\f}{\frac}

\title[DNS of a self-similar APG TBL at the verge of separation]
{Direct numerical simulation of a self-similar adverse pressure gradient turbulent boundary layer at the verge of separation}

\author[V.~Kitsios et al.]
{V.~Kitsios$^{1,2}$, A.~Sekimoto$^1$, C.~Atkinson$^1$, J.A.~Sillero$^3$, G.~Borrell$^3$, A.G.~Gungor$^4$, J.~Jim\'enez$^3$ and J.~Soria$^{1,5}$}

\affiliation{
$^1$Laboratory For Turbulence Research in Aerospace and Combustion, 
Department of Mechanical and Aerospace Engineering, 
Monash University, Clayton 3800, AUSTRALIA \\ [\affilskip]
$^2$CSIRO Oceans and Atmosphere, Hobart 3700, AUSTRALIA \\ [\affilskip]
$^3$ School of Aeronautics, Universidad Polit\'ecnica de Madrid, Pza. \\
Cardenal Cisneros 3, E-28040 Madrid, SPAIN \\ [\affilskip]
$^4$ Istanbul Technical University, \\
Department of Astronautical Engineering, Maslak 34469  Istanbul, TURKEY \\ [\affilskip]
$^5$Department of Aeronautical Engineering, King Abdulaziz University, \\
Jeddah 21589, Kingdom of Saudi Arabia
}

\pubyear{??}
\volume{??}
\pagerange{??}
\date{15 May 2015 and in revised form \today}

\begin{document}

\maketitle

\begin{abstract}
The statistical properties are presented for the direct numerical simulation (DNS) of a self-similar adverse pressure gradient (APG) turbulent boundary layer (TBL) at the verge of separation.
The APG TBL has a momentum thickness based Reynolds number range from $Re_{\delta_2}=570$ to $13800$, with a self-similar region from $Re_{\delta_2} = 10000$ to $12300$.
Within this domain the average non-dimensional pressure gradient parameter $\beta=39$, where for a unit density $\beta=\delta_1 P_e^\prime / \tau_w$, with $\delta_1$ the displacement thickness, $\tau_w$ the mean shear stress at the wall, and $P_e^\prime$ the farfield pressure gradient.
This flow is compared to previous zero pressure gradient (ZPG) and mild APG TBL ($\beta=1$) results of similar Reynolds number.
All flows are generated via the DNS of a TBL on a flat surface with farfield boundary conditions tailored to apply the desired pressure gradient.
The conditions for self-similarity, and the appropriate length and velocity scales are derived.
The mean and Reynolds stress profiles are shown to collapse when non-dimensionalised on the basis of these length and velocity scales.
As the pressure gradient increases, the extent of the wake region in the mean streamwise velocity profiles increases, whilst the extent of the log-layer and viscous sub-layer decreases.
The Reynolds stress, production and dissipation profiles of the APG TBL cases exhibit a second outer peak, which becomes more pronounced and more spatially localised with increasing pressure gradient.
This outer peak is located at the point of inflection of the mean velocity profiles, and is suggestive of the presence of a shear flow instability.
The maximum streamwise velocity variance is located at a wall normal position of $\delta_1$ of spanwise wavelength of $2\delta_1$.
In summary as the pressure gradient increases the flow has properties less like a ZPG TBL and more akin to a free shear layer.
\end{abstract}

\section{Introduction}
\label{sec:introduction}

The efficiency of many engineering systems is dependent upon turbulent boundary layers (TBL) remaining attached to convex curved surfaces, and as such operate in regions of adverse pressure gradient (APG).
Adverse pressure gradients are observed in both internal duct flows \citep{Mathis2008_ETFS}, and on external flows such as those over aircraft wings or wind turbine blades \citep{Kitsios2011_JFM}.
Separation of the TBL in these applications can result in sub-optimal performance, and in some cases may result in catastrophic consequences.
The above mentioned flow configurations are difficult to study systematically, since the pressure gradients are continually changing in the direction of the flow.
There has been much theoretical, experimental and numerical research into TBL, the vast majority of which has been focused on the zero pressure gradient (ZPG) case.
Concerning the APG TBL, however, many aspects of the scaling, structure and stability remain unresolved.
The study of canonical APG TBL is, therefore, of utmost importance.

The self-similar APG TBL is arguably the most appropriate canonical form to study.
A TBL (or region thereof) is deemed self-similar if the terms in the governing equations of motion have the same proportionality with streamwise position \citep{Townsend1956_book,Mellor1966_JFM,George1993_inbook}.
\cite{Mellor1966_JFM} developed this idea to determine that in a self-similar TBL the non-dimensional pressure gradient, $\beta=\delta_1 P_e^\prime / \tau_w$, must be independent of the streamwise position, where $\delta_1$ is the displacement thickness, $\tau_w$ is the mean shear stress at the wall, $P_e^\prime$ is the farfield pressure gradient, and we have prescribed a unit density.
This condition for self-similarity, however, will be broadened in section~\ref{sec:similarity}.
The non-dimensional pressure gradient parameter can be used to classify the various types of TBL into: a ZPG TBL of $\beta=0$; favourable pressure gradient (FPG) of $\beta<0$; APG of $\beta>0$; and an APG TBL immediately prior to separation where $\beta\rightarrow \infty$.
Herein lies the importance of the self-similar canonical TBL.
Imagine two TBL: a FPG decelerating to ZPG; and an APG accelerated to ZPG.
The flow structure, statistics, stability properties and scaling at the position of ZPG in these flows are different from each other, and also different from the canonical ZPG flow \citep{Perry2002_JFM}.
The dynamical properties are dependent upon the specific streamwise distribution of the pressure gradient (also referred to as historical effects).
This illustrates the difficulties in studying APG TBL.
The value of considering the self-similar case in particular is that it minimises (if not removes) the impact of such historical effects.

Theoretical studies of the APG TBL have largely concentrated on the self-similar canonical form.
For a given pressure gradient, theoretical work has focused on deriving the conditions for self-similarity, including the appropriate length and velocity scales necessary to collapse statistical profiles at various streamwise positions onto a single set of profiles \citep{Townsend1956_book,Mellor1966_JFM,Mellor1966b_JFM,Durbin1992_JFM,George1993_inbook,Perry1995_JFM,Marusic1995_JFM,Castillo2004_JFE}.
Additional studies have concentrated specifically on the limiting zero mean wall shear stress ($\beta\rightarrow \infty$) self-similar APG TBL, which is the scenario immediately prior to the point of mean separation \citep{Townsend1960_JFM,Chawla1973_IJES}.
\cite{Skote2002_JFM} propose an alternate viscous velocity scale, based on the pressure gradient, that is finite and non-zero when $\beta\rightarrow\infty$ for the incipiently separated APG TBL.
\cite{Zagarola1998_JFM,Nickels2004_JFM,Maciel2006_AIAA} also attempted to collapse the statistical profiles of non-self-similar APG TBL using various definitions of the pertinent velocity and length scales.

Several experimental campaigns have been undertaken to study the effect of pressure gradients in non-self-similar APG TBLs.
The early studies focused on one-point statistics \citep{Simpson1977_JFM,Cutler1989_JFM,Elsberry2000_JFM,Aubertine2005_JFM,Monty2011_IJHFF}, with more recent measurements elucidating the streamwise structure of such flows \citep{Rahgozar2011_ETFS}.
A smaller number of self-similar TBL experiments have been undertaken, in which the statistical profiles at various streamwise positions collapse under the appropriate scaling \citep{Stratford1959_JFM,Skare1994_JFM,Atkinson2015_FSSIC}.
The limiting separation case was studied in \cite{Skare1994_JFM}, of maximum $\beta=21.4$, with a momentum thickness based Reynolds number $Re_{\delta_2}=5.4\times10^4$.
Across all of the experimental studies, a second outer peak is observed in the variance of the velocity fluctuations, located further away from the wall than the inner peak of the ZPG TBL.
This outer peak also becomes more prominent with increasing pressure gradient.

Direct numerical simulations have also been undertaken of both self-similar and non-self-similar APG TBLs.
The following DNS are all performed in rectangular domains, with the APG applied via a prescribed farfield boundary condition (BC).
\cite{Spalart1993_JFM} produced the first APG TBL DNS, producing a non-self-similar attached TBL of maximum $Re_{\delta_2}=1600$ and $\beta=2$.
DNS of separated APG flows include the studies of \cite{Na1998_JFM,Chong1998_JFM,Skote2002_JFM} and \cite{Gungor2012_JT,Gungor2016_IJHFF}, with the latter study having the largest Reynolds number of $Re_{\delta_2}=2175$.
There have also been various DNS of self-similar APG TBL attempted \citep{Skote1998_FTC,Lee2008_IJHFF,Kitsios2016_IJHFF}.
Two DNS were presented in \cite{Skote1998_FTC}: the first with Reynolds number range $Re_{\delta_2}=390$ to $620$ and $\beta=0.24$; and the second of range $Re_{\delta_2}=430$ to $690$ with $\beta=0.65$.
In the study of \cite{Lee2008_IJHFF} the APG TBL DNS has a Reynolds number range of $Re_{\delta_2}=1200$ to $1400$, and $\beta=1.68$.
In the most recent simulation of \cite{Kitsios2016_IJHFF} an APG TBL DNS was undertaken with a Reynolds number range of $Re_{\delta_2}=300$ to $6000$.
They demonstrated self-similarity of the TBL from $Re_{\delta_2}=3500$ to $4800$, within which $\beta=1$.

The present study will add to the current body of APG TBL DNS databases, in particular addressing the need for high Reynolds number high pressure gradient self-similar flows.
We present a DNS of an APG TBL with a momentum thickness based Reynolds number range from $Re_{\delta_2}=570$ to $13800$ (of equivalent displacement thickness based Reynolds number range from $Re_{\delta_1}=1110$ to $31500$), with a self-similar region spanning a Reynolds numbers $Re_{\delta_2} = 10000$ to $12300$ (or $Re_{\delta_1} = 22200$ to $28800$).
This is larger in both Reynolds number range and magnitude than the aforementioned APG TBL DNS studies.
Within the self-similar region the average pressure gradient parameter $\beta=39$.
The analysis to follow focuses on properties that describe and explain the physics principally in the outer part of the flow.

The manuscript is organised as follows.
Firstly in section~\ref{sec:dns}, an overview of the TBL DNS code is presented along with the farfield BC required to generate the self-similar APG TBL.
The APG TBL is next characterised in section~\ref{sec:flow_characterisation} and compared to the reference ZPG TBL ($\beta=0$) and mild APG TBL ($\beta=1$) of \cite{Kitsios2016_IJHFF}, on the basis of typical boundary layer properties.
In section~\ref{sec:similarity}, the conditions for self-similarity (and associated scaling) are derived from the boundary layer equations and evaluated for each of the TBL.
In section~\ref{sec:mean_profiles} the degree of self-similarity of the strong APG TBL is assessed by comparing mean streamwise velocity profiles across various streamwise stations.
The impact of increasing pressure gradient on the wake, log-layer, viscous sub-layer, and inflection points are also presented.
In section~\ref{sec:Reynolds_stresses} self-similarity is again assessed by comparing profiles of Reynolds stress at various locations.
The influence of the pressure gradient on the existence, location and magnitude of the inner and outer Reynolds stresses peaks is also discussed.
A physical model is proposed that explains how the generation of the turbulent fluctuations changes with pressure gradient.
In section~\ref{sec:momentum_terms} wall normal profiles of the boundary layer momentum terms quantify how the relative magnitude of each of these terms change with increasing pressure gradient, and identify the direction of momentum transfer between the mean and fluctuating fields.
The turbulent kinetic energy budgets illustrate the sources, sinks and transfers of these turbulent fluctuations in section~\ref{sec:budget_profiles}.
The wall normal location and spanwise scale that contribute the most to the total fluctuations in the outer region are determined from the streamwise velocity spectra in section~\ref{sec:spanwise_spectra}.
At this wall normal location, two-point correlations in section~\ref{sec:two-point_correlations} indicate how the structures become more compact as the pressure gradient increases.
Concluding remarks are made in section~\ref{sec:conclusion}.

\section{Direct numerical simulation}
\label{sec:dns}

In the following sections we present: the algorithmic details of the DNS; boundary conditions necessary to implement the strong APG TBL; definitions of appropriate velocity and integral length scales; and numerical details of the simulations.

\subsection{Algorithmic details}

The code adopted within solves the Navier-Stokes equations in a three-dimensional rectangular volume, with constant density (here set to one) and kinematic viscosity ($\nu$).
The three flow directions are the streamwise ($x$), wall normal ($y$) and spanwise ($z$), with instantaneous velocity components in these directions denoted by $U$, $V$ and $W$, respectively.
Notation used for the derivative operators in these directions are $\p_x \equiv \p / \p x$, $\p_y \equiv \p / \p y$, and $\p_z \equiv \p / \p z$.
Throughout the paper the mean velocity components are represented by $(\langle U \rangle,\langle V \rangle,\langle W \rangle)$, with the averaging undertaken both in time and along the spanwise direction.
The associated fluctuating velocity components are $(u,v,w)$.

Details of the algorithmic approach to solve the equations of motion are as follows.
A fractional-step method is used to solve the governing equations for the velocity and pressure ($P$) fields \citep{Harrow1965_PoF,Perot1993_JCP}.
The grid is staggered in the streamwise and wall normal directions but not in the spanwise.
Fourier decomposition is used in the periodic spanwise direction, with compact finite difference in the aperiodic wall normal and streamwise directions \citep{Lele1992_JCP}.
The equations are stepped forward in time using a modified three sub-step Runge-Kutta scheme \citep{Simens2009_JCP}.
The code utilises MPI and openMP parallelisation to decompose the domain.
For further details on the code and parallelisation, the interested reader should refer to \cite{Borrell2012_CF} and \cite{Sillero2014_PhDthesis}.

\subsection{Boundary conditions}
\label{sec:BCs}

In all TBL the bottom surface is a flat plate with a no-slip (zero velocity) BC, and the spanwise boundaries are periodic.
The following boundary conditions pertain to the strong APG TBL DNS.
Refer to \cite{Kitsios2016_IJHFF} for details of the boundary conditions applied in the mild APG and ZPG TBL DNS. 

\begin{figure*}
\psfrag{ZPG BC smoothed into APG BC}[c][t][1.0]{\rm \small ZPG BC smoothed into APG BC $ \ \ \ \ \ \ \ \ \ \ $}
\psfrag{recycle plane}[c][b][1.0]{\rm \small recycle plane}
\psfrag{ZPG}[c][][1.0]{\rm \small ZPG}
\psfrag{APG}[c][][1.0]{\rm \small $\beta=1$}
\psfrag{ZPG1}[c][][1.0]{$ \ \ $ \rm \small ZPG}
\psfrag{zone1}[c][][1.0]{$ \ \ $ \rm \small zone}
\psfrag{APG2}[c][][1.0]{\rm \small APG}
\psfrag{zone2}[c][][1.0]{\rm \small zone}
\psfrag{acceleration3}[c][][1.0]{$ \ \ \ \ \ \ $ \rm \small acceleration}
\psfrag{zone3}[c][][1.0]{$ \ \ \ \ \ \ \ \ \ \ \ \ \ \ \ \ \ \ $ \rm \small zone}
\psfrag{acceleration}[c][][1.0]{\rm \small acceleration}
\psfrag{zone}[c][][1.0]{\rm \small zone}
\psfrag{recycle}[c][][1.0]{\rm \small recycle}
\psfrag{plane}[c][][1.0]{\rm \small plane}
\psfrag{x/delta1}[c][][1.0]{$x/\delta_1(x_I)$}
\psfrag{vBC}[c][b][1.0]{$V_\infty(x)/U_{e}(x_I)$}
\psfrag{a}[c][t][1.0]{$x_S \ \ $}
\psfrag{b}[c][][1.0]{$ \ \ \ x_F$}
\psfrag{c}[c][t][1.0]{$ \ x_B$}
\psfrag{d}[c][][1.0]{$x_I$}
\psfrag{r}[c][][1.0]{$x_R$}
\psfrag{x0}[c][][1.0]{$ \ x_I$}
\psfrag{xB}[c][][1.0]{$ \ x_B$}
\psfrag{xR}[c][][1.0]{$ \ x_R$}
\begin{center}
  \begin{tabular}{c}
  \\
  \includegraphics[trim = 0mm 25mm 0mm 19mm, clip, width=\textwidth]{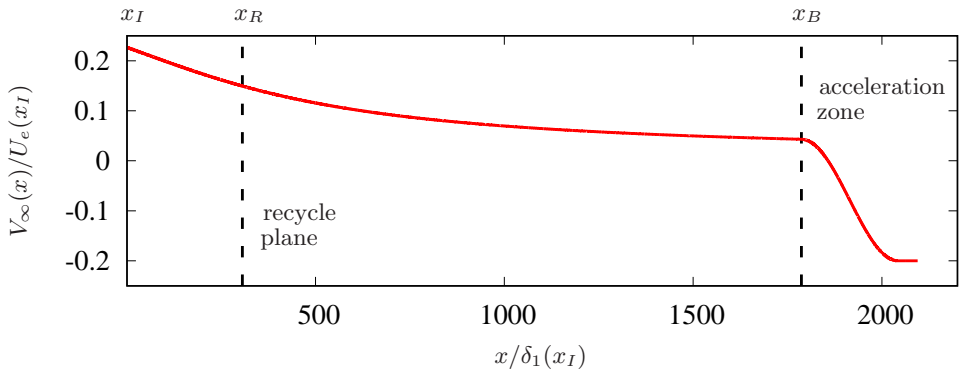} \\
  \\
  \end{tabular}
\end{center}
\caption{Farfield wall normal velocity boundary condition of the strong APG TBL DNS, with
$x_I$ the location of the inlet plane,
$x_R$ the location of the recycling plane, and
$x_B$ the position at which blowing into the computational domain is initiated.
}
\label{fig:BC}
\end{figure*}

Due to the TBL growing in height as it develops in the streamwise direction, at a downstream recycling position a spanwise/wall normal plane is copied and mapped to the inlet BC.
We use a modified version of the recycling method presented in \cite{Sillero2013_PF}, which scales and regrids the instantaneous velocity profiles at the recycling plane to ensure that its reference velocity and length scales match those prescribed at the inlet.
As illustrated in figure~\ref{fig:BC}, the recycling plane is located at $x_R=307\delta_1(x_I)$, where $\delta_1(x_I)$ is the displacement thickness at the inlet of streamwise position $x_I$.
For the purposes of this BC, the reference velocity scale at the recycling plane located at $x_R$ is denoted by $U_R$, and defined as the maximum mean streamwise velocity in the wall normal direction of position $y = \delta(x_R)$.
The associated reference length scale
\begin{eqnarray}
\label{eq:Lref}
  L_R &=& \int_0^{\delta(x_R)} \left( 1 - \f{\langle U \rangle(x_R,y)}{U_R} \right) \f{\langle U \rangle(x_R,y)}{U_R} \ dy \ \mbox{,}
\end{eqnarray}
\noindent is defined in a manner analogous to the classical momentum thickness.
Throughout the paper, $\delta$, denotes the point of maximum streamwise velocity along the profile.
The spanwise homogeneous Fourier mode of the initial inlet profiles are rescaled and interpolated from the time averaged profiles of a previous preliminary simulation.
These time averaged profiles, were selected at the streamwise position of the preliminary simulation with a shape factor of $H=2.35$, which is the empirical value for an incipient APG TBL from the study of \cite{Mellor1966_JFM}.

At the farfield boundary a zero spanwise vorticity condition is applied, and the wall normal velocity specified.
It is important that the wall normal velocity be prescribed, as opposed to the streamwise velocity, so as not to over constrain the system \citep{Rheinboldt1956_ZAMM}.
The wall normal velocity at the farfield boundary is based on the potential flow solution in an expanding duct, corrected for the growth of the boundary layer.
The general potential flow solution is first derived, followed by the necessary modifications to account for the boundary layer growth.
According to \cite{Mellor1966_JFM} for the case of incipient separation, the outer reference velocity must be proportional to $(x-x_0)^{m}$, where $x_0$ is the virtual origin of the boundary layer, and the exponent $m=-0.23$.
The general potential flow solution of an expanding duct that produces this functional form along the centreline of the duct is given by the streamfunction
\begin{eqnarray}
\label{eq:streamfunction}
  \psi_{PF}(\hat{x},\hat{y}) &=& A r^{m+1} \cos(\gamma) \ \mbox{, where} \\
\label{eq:r}
  r^2 &=& \hat{x}^2 + \hat{y}^2 \ \mbox{, } \\
\label{eq:gamma}
  \gamma &=& (m+1)\arctan(\hat{y}/\hat{x}) \ \mbox{,}
\end{eqnarray}  
\noindent and the constant $A$ is a scaling parameter, with $\hat{x}$ and $\hat{y}$ the streamwise and wall normal coordinates, respectively.
The general potential flow streamwise and wall normal velocity components can be calculated from the streamfunction, by
\begin{eqnarray}
\label{eq:uAPG}
  U_{PF}(\hat{x},\hat{y}) &=& \p_y \psi_{PF} = A (m+1) \left[ \hat{x} r^{m-1} \cos(\gamma) + \hat{y} r^{m-1} \sin(\gamma) \right] \ \mbox{, and} \\
\label{eq:vAPG}
  V_{PF}(\hat{x},\hat{y}) &=& -\p_x \psi_{PF} = A (m+1) \left[ \hat{y} r^{m-1} \cos(\gamma) - \hat{x} r^{m-1} \sin(\gamma) \right] \ \mbox{.} 
\end{eqnarray}
\noindent Note at the centreline of the expanding duct $\hat{y}=0$, which means $r=\hat{x}$ and $\gamma=0$.
When substituted into (\ref{eq:uAPG}), $U_{PF}(\hat{x},0)=A(m+1)\hat{x}^m$, which has the proportionality with streamwise position as specified in \cite{Mellor1966_JFM}.
As previously intimated, this potential flow solution does not account for boundary layer growth.
The displacement thickness of the self-similar TBL at the verge of separation grows linearly, with the functional form $\delta_1(x) = K(x-x_0)$, where $K=0.041$ \citep{Mellor1966_JFM}.
To ensure the correct streamwise velocity decay along the displacement thickness height, the relationship between the general potential flow coordinates ($\hat{x}$,$\hat{y}$) and the coordinates of the DNS ($x,y$) is required to be
\begin{eqnarray}
\label{eq:xHat}
  \hat{x} &=& x-x_0 \ \mbox{, and} \\
\label{eq:yHat}
  \hat{y} &=& y-K(x-x_0) \ \mbox{.} 
\end{eqnarray}
\noindent By substituting the above relationships into (\ref{eq:uAPG}), it can be shown that $U_{PF}(x-x_0,y-K(x-x_0))$ now has the correct decay of the streamwise velocity along the displacement thickness height.

To finalise this farfield suction BC two parameters must be determined: the scale factor $A$; and the virtual origin of the boundary layer $x_0$.
The parameter $A$ is calculated such that the modified potential flow solution $U_{PF}(x-x_0,y-K(x-x_0))$ from (\ref{eq:uAPG}), matches the prescribed inlet streamwise velocity profile at the boundary layer edge.
The virtual origin of the boundary layer ($x_0$) is calculated by extending back the streamline from the boundary layer edge of the inlet profile, of position $(x,y)=(x_I,\delta(x_I))$, to give $x_0=x_I-\delta(x_I) \times U(x_I,\delta(x_I))/V(x_I,\delta(x_I))$.
Finally, the farfield wall normal BC is given by 
\begin{eqnarray}
  V_\infty(x) &=& V_{PF}(x-x_0,y_{BC}-K(x-x_0)) \ \mbox{, } 
\end{eqnarray}
\noindent along the length of the domain, where $y_{BC}$ is the wall normal position of the top boundary. 
This farfield BC is transitioned from suction ($V_\infty(x)>0$, fluid leaving the computational domain) at $x_B=1790\delta_1(x_I)$ to blowing ($V_\infty(x)<0$, fluid entering the computational domain) at the outflow.
This reduces the number of instantaneous reversed flow events at the downstream outflow boundary, and helps to ensure numerical stability.
The farfield boundary condition, $V_{\infty}(x)/U_{e}(x_I)$, is illustrated in figure~\ref{fig:BC}, where $U_{e}(x_I)$ is the reference streamwise velocity at the inlet defined in (\ref{eq:Ue}) of the following section.
Note the outer portion of the inlet profile for $y>\delta(x_I)$ is defined by the same potential flow solution, specifically $U_{PF}(x_I-x_0,y-K(x_I-x_0))$, which ensures consistency with the application of $V_\infty(x)$.

\subsection{Definition of velocity and integral length scales}

At the farfield wall normal boundary of a strong APG TBL, $\p V_\infty/\p x$ is significant and negative.
This means that for the farfield to have zero spanwise vorticity, $\p \langle U \rangle /\p y$ at the boundary must also be less than zero.
The $\langle U \rangle$ profile, therefore, has a maximum in $y$.
In this case, the classical definitions of $\delta_1$ and $\delta_2$, are not appropriate since the velocity profiles do not approach a constant value.
Due to these properties, for the presentation of the results within, we adopt the definitions of reference velocity ($U_e$), displacement thickness ($\delta_1$), and momentum thickness ($\delta_2$) akin to that of \cite{Spalart1993_JFM}.
This reference velocity scale, as first proposed in \cite{Lighthill1963_chapter}, is given by
\begin{eqnarray}
\label{eq:Ue}
  U_e(x) &=& U_\Omega(x,y_\Omega) \ \mbox{, where} \\
\label{eq:Uomega}
  U_\Omega(x,y) &=& -\int_0^{y} \langle \Omega_z \rangle (x,\tilde{y}) \ d\tilde{y} \ \mbox{,}
\end{eqnarray}
\noindent with $\langle \Omega_z \rangle$ the mean spanwise vorticity, and $y_\Omega$ is the wall normal position at which $\langle \Omega_z \rangle$ is $0.2\%$ of the mean vorticity at the wall.
The integral length scales are given by
\begin{eqnarray}
\label{eq:delta1_def}
  \delta_1(x) &=& \f{-1}{U_e}  \int_0^{y_\Omega} y \langle\Omega_z\rangle(x,y) \ dy \ \mbox{, and} \\
\label{eq:delta2_def}
  \delta_2(x) &=& \f{-2}{U_e^2} \int_0^{y_\Omega} y U_\Omega \langle\Omega_z\rangle(x,y) \ dy - \delta_1(x) \ \mbox{.}
\end{eqnarray}

\subsection{Numerical details}
\label{sec:numerical_details}

The numerical details of the present simulations are summarised in table~\ref{table:numerics}.
The table lists the number of collocation points in the streamwise ($N_x$) and wall normal ($N_y$) directions, and the number of spanwise Fourier modes after de-aliasing ($N_z$).
The extent of the computational domain in the streamwise, wall normal and spanwise directions is denoted by $L_x$, $L_y$, and $L_z$, respectively.
The computational domain size is non-dimensionalised with respect to the displacement thickness ($\delta_{1}(x_\star)$), at the streamwise position, $x_\star$, which is where the displacement thickness based Reynolds number $Re_{\delta_1}\equiv U_e \delta_1(x_\star) / \nu=4800$.
The streamwise dependent boundary layer properties for each of the three TBLs are later presented starting from $x_\star$, and hence the same Reynolds number.
The strong APG TBL DNS has a larger wall normal domain ($L_y$) and more points in this direction ($N_y$), than the mild APG TBL, which in turn has a larger wall normal domain than the ZPG TBL simulations.
This is necessary since the present APG TBL expands more quickly while evolving in the streamwise direction.
Two additional strong APG TBL DNS were also undertaken: one with $N_y=700$ and $L_y$ at $58\%$ of the wall normal domain of the present strong APG TBL DNS; and a second with $N_y=900$ and $L_y$ at $78\%$ of the present strong APG TBL DNS wall normal domain.
It was found that a wall normal domain of the size of the present strong APG TBL DNS was required in order for the potential flow farfield BC to be applied at a location of sufficiently low mean spanwise vorticity.

The grid resolutions are also presented in table~\ref{table:numerics}.
The grid spacings in the streamwise ($\Delta x$) and spanwise directions ($\Delta z$) are constant.
The smallest wall normal grid spacing is located at the wall ($\Delta y_{wall}$), and increases monotonically to the maximum wall normal grid spacing located at the farfield boundary ($\Delta y_{\infty}$).
These grid spacings are again non-dimensionalised by $\delta_{1}(x_\star)$.
Using figure~\ref{fig:BL_props}(b) one can determine the relative resolutions and domain sizes with respect to the displacement thicknesses at other streamwise positions.
The relative size of the boundary layers at $Re_{\delta_1}=4800$ is also presented in table~\ref{table:numerics} by listing the ratio of $\delta_{1}(x_\star)$ for each TBL to that of the strong APG TBL.
The Courant number is set to unity.
The time ($T$) taken to accumulate the statistics in terms of the eddy-turnover times at reference positions $x_\star$ and $x_{DoI}$ are also listed in table~\ref{table:numerics}.

\begin{table}
\caption{Numerical details of the ZPG, mild APG and strong APG TBL DNS:
number of collocation points in the streamwise ($N_x$) and wall normal ($N_y$) directions, and the number of spanwise Fourier modes after de-aliasing ($N_z$);
domain size $L_x$, $L_y$ and $L_z$ in these respective directions non-dimensionalised by the displacement layer thickness ($\delta_{1}$) at the position, $x_{\star}$, where $Re_{\delta_1}=4800$;
uniform streamwise ($\Delta x$) and spanwise ($\Delta z$) grid spacing and wall normal grid spacing at the wall ($\Delta y_{\rm wall}$) and at the farfield boundary ($\Delta y_{\infty}$) non-dimensionalised by $\delta_{1}(x_{\star})$;
$\delta_{1}(x_{\star})$ relative to $\delta_{1}(x_{\star})$ of the strong APG TBL;
$Re_{\delta_1}$ and $Re_{\delta_2}$ range of the domain of interest (DoI);
streamwise extent of the domain of interest ($L_{DoI}$) in terms of $\delta_1(x_{\star})$, and the displacement thickness at the beginning of the domain of interest $\delta_1(x_{DoI})$; and
the time taken to accumulate the statistics ($T$) in terms of the eddy-turnover times at $x_\star$ (i.e. $T U_\infty(x_\star) / \delta_1(x_\star)$) and at $x_{DoI}$ (i.e. $T U_\infty(x_{DoI}) / \delta_1(x_{DoI})$).
}
\centering
\begin{tabular}{cccc}
  & & & \\
  \hline
  & $ \ \ \ \ \ \ \ \ $ ZPG $ \ \ \ \ \ \ \ \ $ & $ \ \ \ \ \ $ Mild APG $ \ \ \ \ \ $ & $ \ \ \ $ Strong APG $ \ \ \ $ \\
  \hline
  nominal $\beta$ & $0$ & $1$ & $39$ \\ 
  $N_x$ & $8193$ & $8193$ & $8193$ \\
  $N_y$ & $315$ & $500$ & $1000$ \\ 
  $N_z$ & $1362$ & $1362$ & $1362$ \\
  $L_x/\delta_{1}(x_{\star})$ & $480$ & $345$ & $303$ \\
  $L_y/\delta_{1}(x_{\star})$ & $22.7$ & $29.8$ & $73.4$ \\
  $L_z/\delta_{1}(x_{\star})$ & $80.1$ & $57.6$ & $50.7$ \\
  $\Delta x/\delta_{1}(x_{\star})$ & $0.0585$ & $0.0421$ & $0.0370$ \\
  $\Delta y_{\rm wall}/\delta_{1}(x_{\star})$ & $1.53\times10^{-3}$ & $1.10\times10^{-3}$ & $9.71\times10^{-4}$ \\
  $\Delta y_{\infty}/\delta_{1}(x_{\star})$ & $0.0992$ & $0.0714$ & $0.254$ \\
  $\Delta z/\delta_{1}(x_{\star})$ & $0.0585$ & $0.0421$ & $0.0370$ \\
  $\delta_{1}(x_{\star})/\delta_{1}(x_\star\mbox{ ; Strong APG})$ & $0.63$ & $0.88$ & $1$ \\
  $Re_{\delta_1}$ range in DoI & $4800 \rightarrow 5280$ & $4800 \rightarrow 5280$ & $22200 \rightarrow 28800$ \\
  $Re_{\delta_2}$ range in DoI & $3500 \rightarrow 3880$ & $3100 \rightarrow 3440$ & $10000 \rightarrow 12300$ \\  
  $L_{DoI}/\delta_1(x_{\star})$ & $82$ & $20$ & $37$ \\
  $L_{DoI}/\delta_1(x_{DoI})$ & $82$ & $20$ & $7$ \\
  $T U_e(x_\star) / \delta_1(x_\star)$ & $621$ & $720$ & $1160$ \\
  $T U_e(x_{DoI}) / \delta_1(x_{DoI})$ & $621$ & $720$ & $165$ \\
  \hline
\end{tabular}
\label{table:numerics}
\end{table}

\section{Flow characterisation}
\label{sec:flow_characterisation}

\begin{figure}
\begin{tabular}{c}
    \includegraphics[trim = 0mm 0mm 0mm 0mm, clip, width=\textwidth]{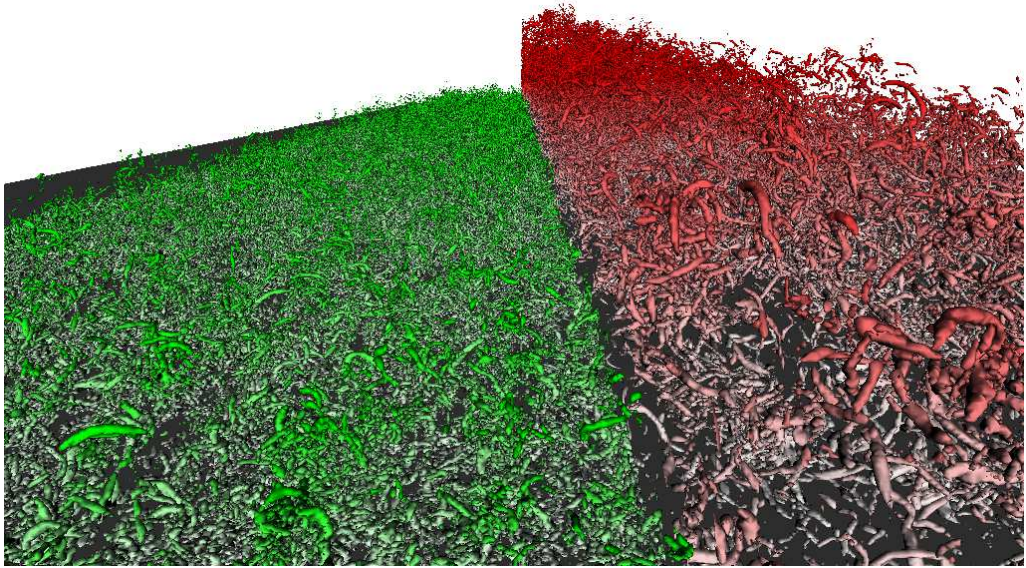} \\
\end{tabular}
\caption{Instantaneous iso-surfaces of the discriminant ($D$) of the velocity gradient tensor for the:
ZPG TBL (left, green) with iso-surface levels of $D / \langle D \rangle_{yz} = 1$; and
strong APG TBL (right, red) with iso-surface levels of $D / \langle D \rangle_{yz} = 10$, where $\langle D \rangle_{yz} (x)$ is the discriminant averaged within the local boundary layer height ($\delta(x)$) and over the span.
The streamwise direction is into the page, the wall normal direction is normal to the dark grey surface, and the spanwise direction is across the image from left to right.
The colour intensity of the iso-surfaces increases with distance from the wall.
}
\label{fig:structures}
\end{figure}

To give a qualitative indication of the differences in the size and complexity of the boundary layers, figure~\ref{fig:structures} illustrates instantaneous iso-surfaces of the discriminant of the velocity gradient tensor ($D$) for the ZPG TBL (left, green) and the strong APG TBL (right, red). 
Quantitative comparisons between these flows and also the mild APG case are to follow.
In figure~\ref{fig:structures}, the streamwise direction is into the page, the wall normal direction is normal to the dark grey surface, and the spanwise direction is across the image from left to right.
The computational domain appears to shrink in the spanwise direction due to perspective foreshortening.
At the inlet plane the ZPG and APG boundary layers have the same boundary layer thickness and maximum mean streamwise velocity.
The iso-surface levels for the ZPG TBL is $D / \langle D \rangle_{yz} = 1$, where $\langle D \rangle_{yz}$ is the discriminant averaged within the local boundary layer height ($\delta(x)$) and over the span.
The iso-surface levels for the strong APG TBL are $D / \langle D \rangle_{yz} = 10$.
The colour intensity of these iso-surfaces increases with distance from the wall.
This figure clearly illustrates the shear size and complexity of the two flows, with the APG TBL undergoing significantly more wall normal expansion as it progresses in the streamwise direction than its ZPG counterpart.

\begin{figure}
\centering
\psfrag{ReDelta/1000}[c][][1.0]{$Re_{\delta_1}\times10^{-3}$}
\psfrag{ZPG}[c][][1.0]{$\beta=0$}
\psfrag{beta1}[c][][1.0]{$\beta=1$}
\psfrag{betaInf}[c][][1.0]{$\beta=39$}
\psfrag{Cf*1e3}[c][b][1.0]{$C_f(x) \times 10^3$}
\psfrag{tauW*1e3}[c][b][1.0]{$\tau_w(x) \times 10^3$}
\psfrag{uRef}[c][b][1.0]{$U_e(x) \ / \ U_e(x_I)$}
\psfrag{H}[c][b][1.0]{$H(x)$}
\psfrag{beta}[c][b][1.0]{$\beta$}
\psfrag{delta1}[c][b][1.0]{$\delta_1/\delta_1(x_{\star})$}
\psfrag{x/delta(ReDelta=5000)}[c][][1.0]{$(x-x_{\star})/\delta_1(x_{\star})$}
\begin{tabular}{ll}
    & \\
    (a) & (b) \\
    \includegraphics[trim = 3mm 0mm 3mm 2mm, clip, width=0.5\textwidth]{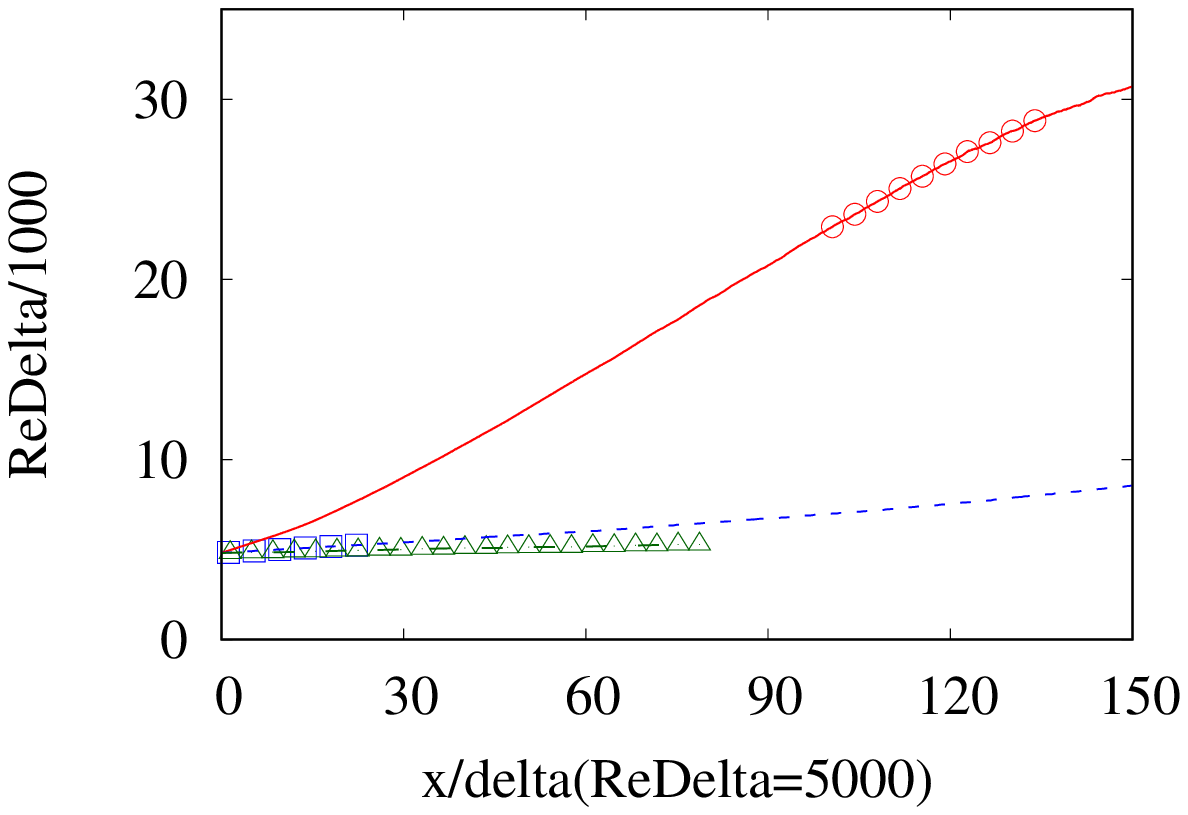} &
    \includegraphics[trim = 3mm 0mm 3mm 2mm, clip, width=0.5\textwidth]{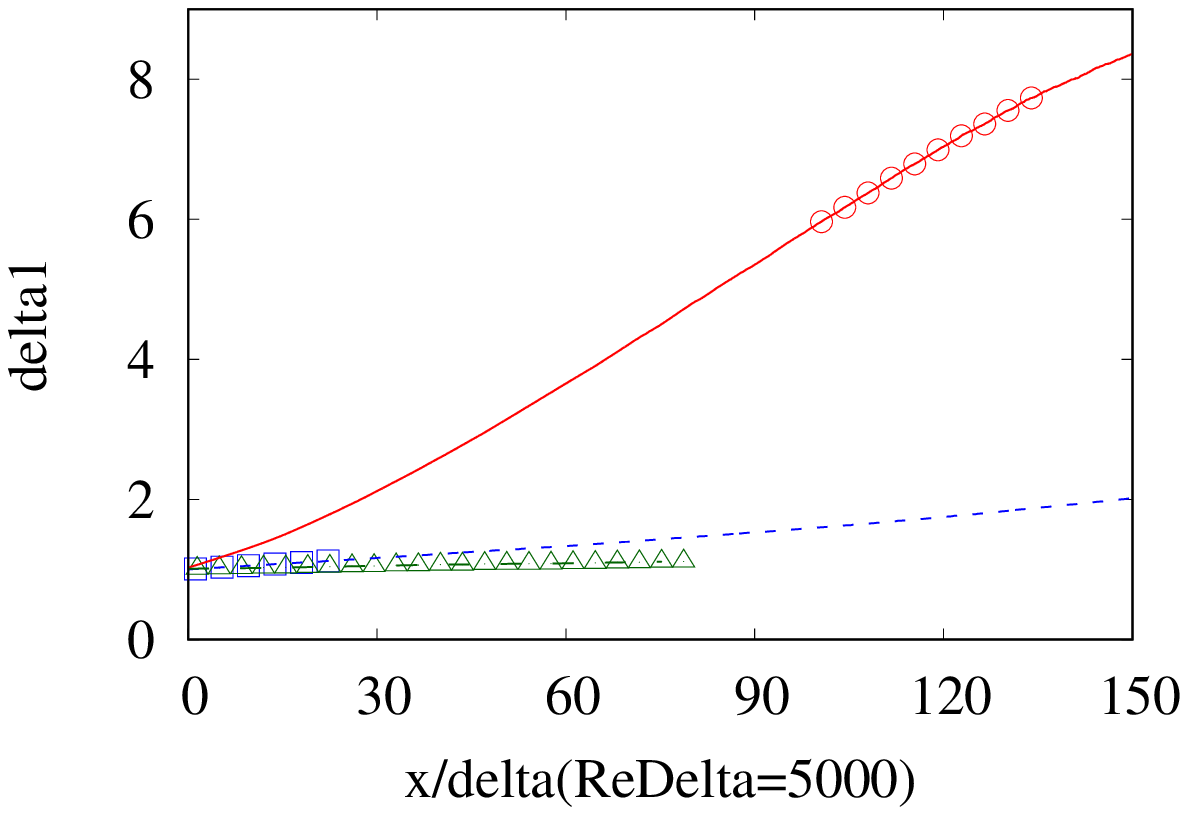} \\
    (c) & (d) \\
    \includegraphics[trim = 3mm 0mm 3mm 2mm, clip, width=0.5\textwidth]{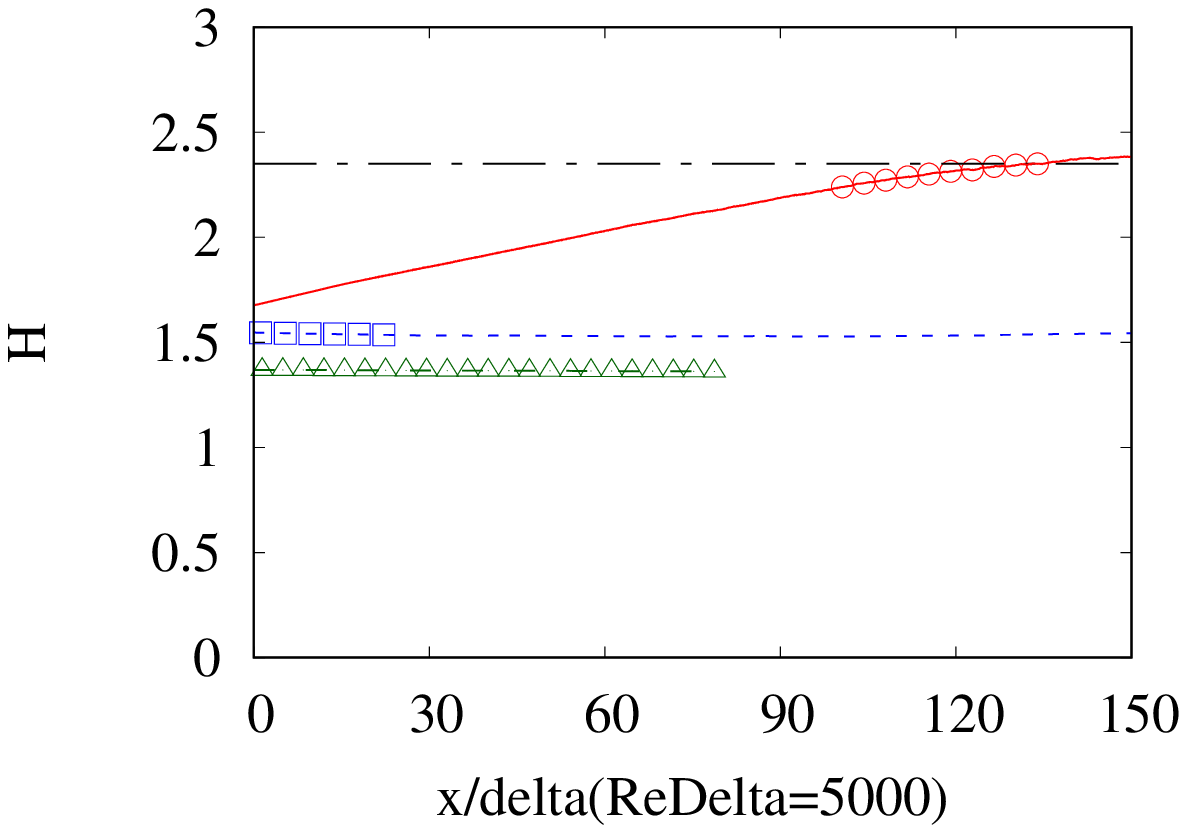} &
    \includegraphics[trim = 3mm 0mm 3mm 2mm, clip, width=0.5\textwidth]{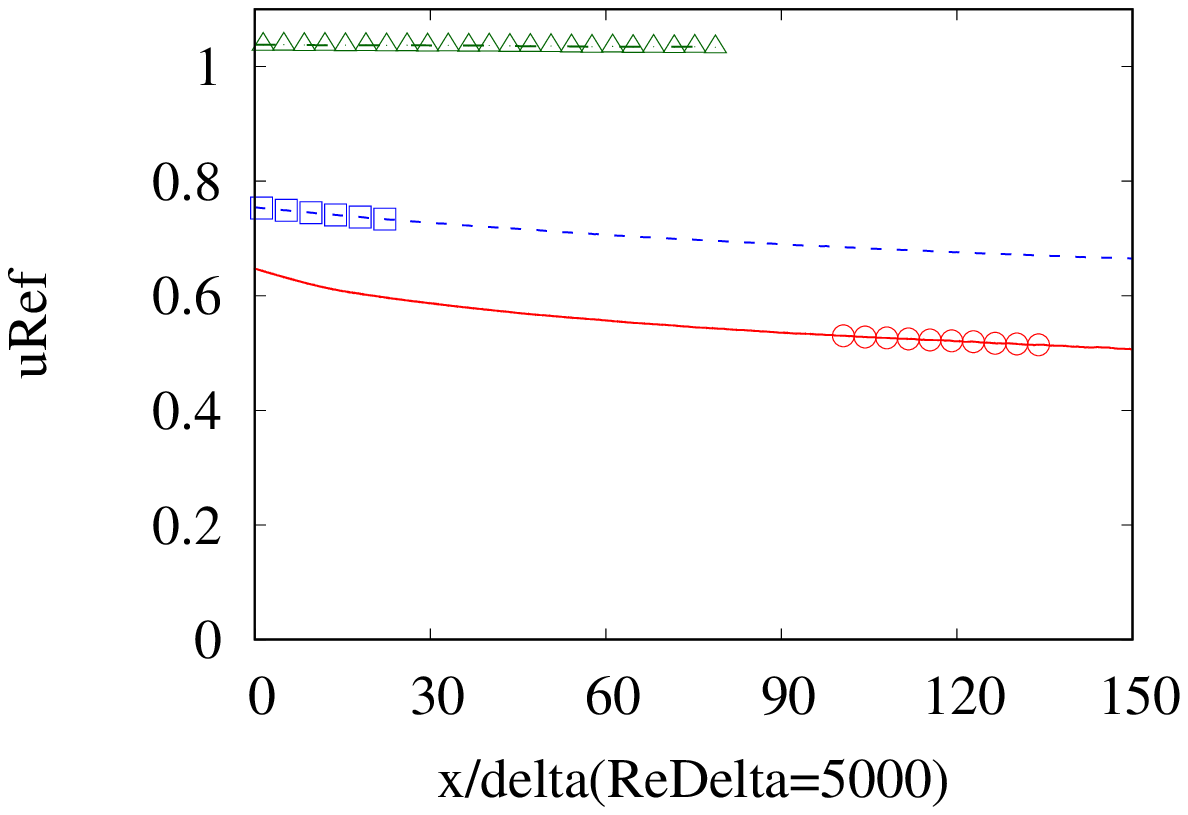} \\
    (e) & (f) \\
    \includegraphics[trim = 3mm 0mm 3mm 2mm, clip, width=0.5\textwidth]{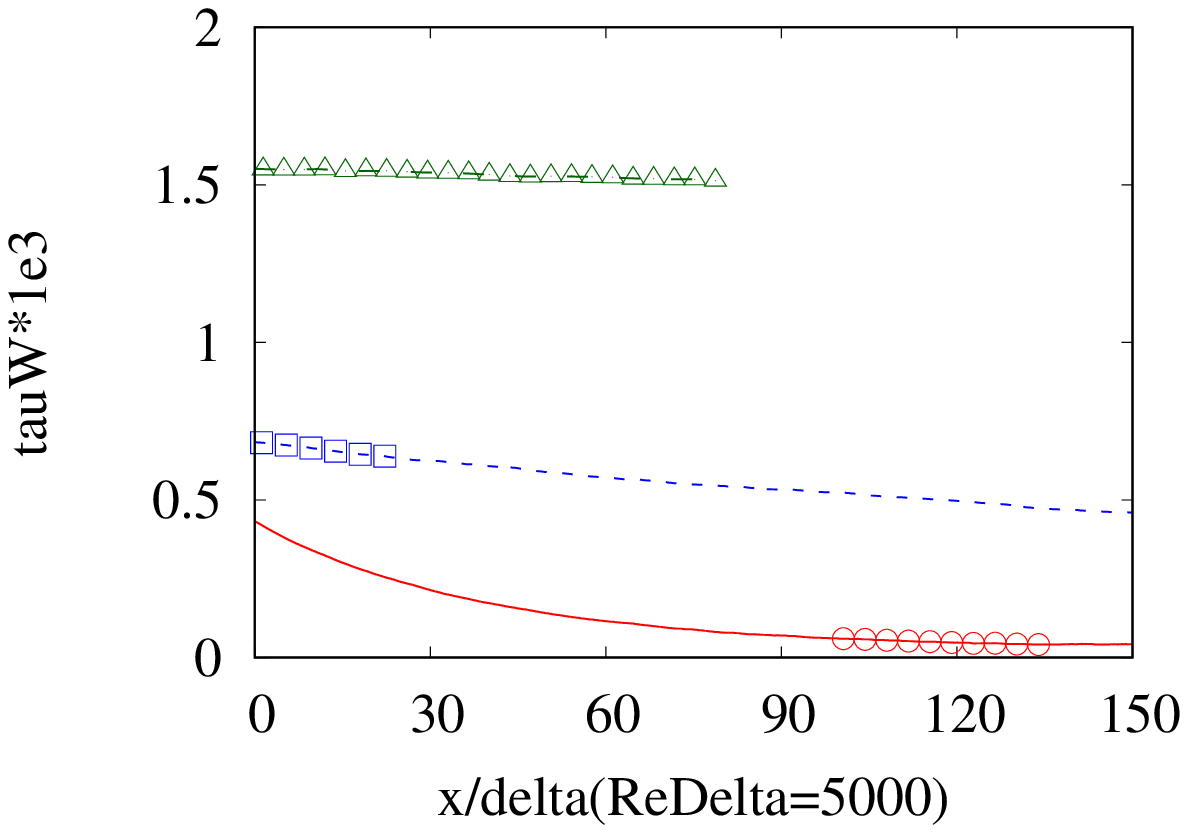} &
    \includegraphics[trim = 3mm 0mm 3mm 2mm, clip, width=0.5\textwidth]{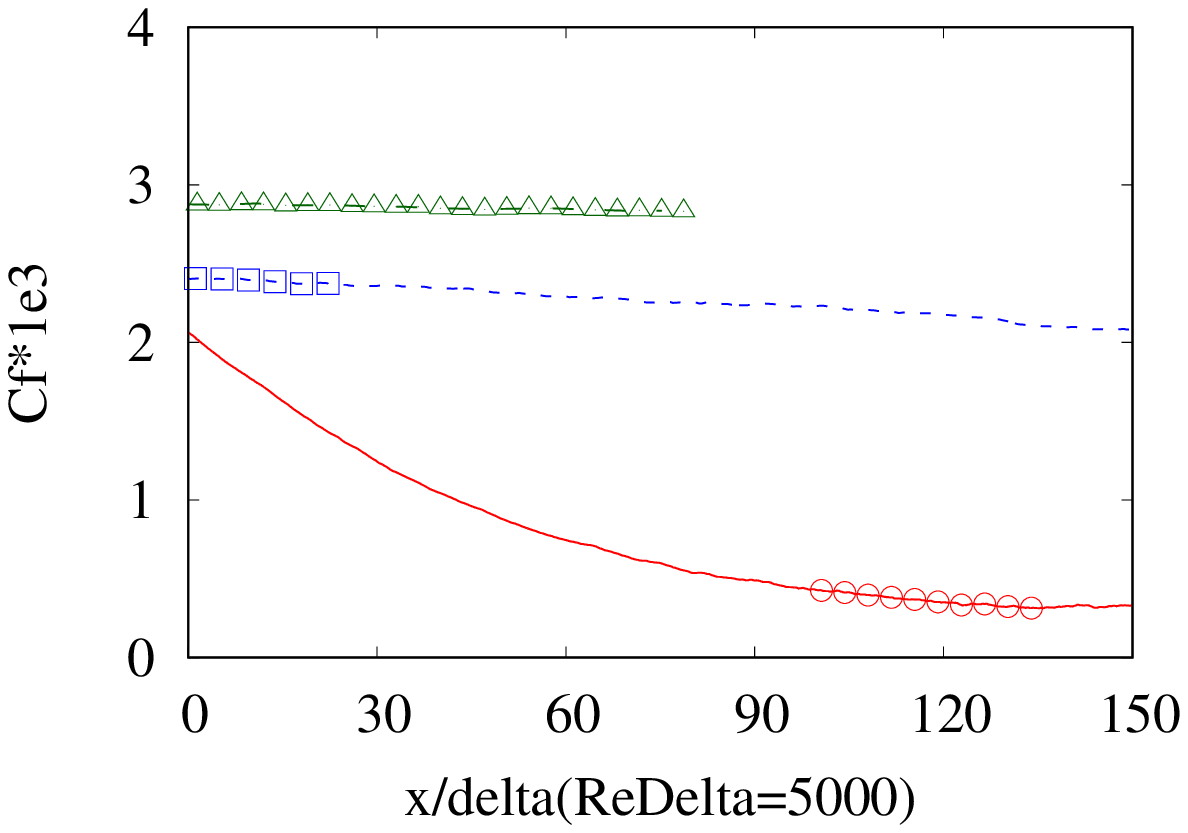} \\
\end{tabular}
\caption{Boundary layer properties of the strong APG (solid red lines, with domain of interest indicated by \textcolor{red}{$\circ$}),
mild APG (short dashed blue lines, with domain of interest indicated by \textcolor{blue}{\opensquare}) and 
ZPG (long dashed green lines, with domain of interest indicated by \textcolor{green}{\opentriangle}) TBL:
(a) displacement thickness based Reynolds number ($Re_{\delta_1}$);
(b) displacement thickness ($\delta_1$);
(c) shape factor, $H=\delta_1/\delta_2$, with empirical value of $H=2.35$ for the incipient case \citep{Mellor1966_JFM} indicated by the black dash-dotted line;
(d) outer reference velocity, $U_e$;
(e) wall shear stress, $\tau_w$; and
(f) skin friction coefficient, $C_f=2\tau_w / U_e^2$.
Note $x_{\star}$ is the streamwise position at which $Re_{\delta_1}=4800$.}
\label{fig:BL_props}
\end{figure}

We now quantitatively show how the boundary layer properties of the ZPG ($\beta=0$), mild APG ($\beta=1$) and the strong APG ($\beta=39$) TBL evolve in the streamwise direction.
Each of the remaining plots in this section, and the plots in section~\ref{sec:similarity}, have $(x-x_{\star})/\delta_1(x_{\star})$ as the independent variable.
Since $x_{\star}$ is the position at which $Re_{\delta_1}=4800$, the shifting of the independent axis by $x_{\star}$ ensures that $Re_{\delta_1}=4800$ at the origin for all of the three TBLs.
The portions of each of the lines in figure~\ref{fig:BL_props} with symbols indicate the respective domains of interest.
For the ZPG case the domain of interest spans $Re_{\delta_1}=4800$ to $5280$, which is in fact the entire illustrated ZPG domain.
The first streamwise position of the domain of interest ($x_{DoI}$) is hence equal to $x_{\star}$.
The mild APG TBL is self-similar over a larger Reynolds number range \citep{Kitsios2016_IJHFF}, however, the domain of interest is purposely selected to span the same $Re_{\delta_1}$ as the ZPG case.
This is done in an attempt to remove any Reynolds number effects and isolate the impact of the pressure gradient.
The strong APG TBL, however, is not in a self-similar state over this same Reynolds number range.
Instead the domain of interest for the strong APG case spans $Re_{\delta_1}=22200$ to $28800$.
The streamwise extent over the respective domains of interest ($L_{DoI}$) is equivalent to $28\delta_1(x_{\star})$ for the ZPG, $20\delta_1(x_{\star})$ for the mild APG TBL, and $37\delta_1(x_{\star})$ (equivalent to $7\delta_1(x_{DoI})$) for the strong APG TBL, as listed in table~\ref{table:numerics}.
The range of the respective domains of interest are also listed in terms of the momentum thickness based Reynolds number, $Re_{\delta_2}$, in table~\ref{table:numerics}.

The displacement thickness Reynolds numbers in figure~\ref{fig:BL_props}(a) illustrate that the domains of interest of the ZPG and mild APG case span the same $Re_{\delta_1}$ range.
Figure~\ref{fig:BL_props}(a) also demonstrates that as the pressure gradient increases, so too does $Re_{\delta_1}\equiv U_e\delta_1/\nu$, due to $\delta_1$ increasing more rapidly with $x$.
The increase in displacement thicknesses ($\delta_1$), as defined in (\ref{eq:delta1_def}), is illustrated in figure~\ref{fig:BL_props}(b).
This length scale is larger in the strong APG TBL compared to the mild APG, which in turn is larger than that of the ZPG TBL.
This indicates that the boundary layer expands in the streamwise direction more rapidly as the pressure gradient increases.
The shape factor $H=\delta_1/\delta_2$ is illustrated in figure~\ref{fig:BL_props}(c), and  is relatively constant over the domain of interest for each case.
The shape factor of the strong APG TBL also approaches the empirical value of $H=2.35$ \citep{Mellor1966_JFM}.
The mild and strong APG TBL are decelerated via the BC as illustrated in figure~\ref{fig:BL_props}(d), where their respective outer reference velocities ($U_e$) decrease with $x$.
The expansion of the boundary layers coincides with a reduction of the mean wall shear stress ($\tau_w$).
In figure~\ref{fig:BL_props}(e), $\tau_w$ decreases with increasing pressure gradient, since the pressure gradient expands the TBL, thus reducing the mean gradient at the wall.
The effect of the boundary layer expansion is also evident in the reduced skin friction coefficient ($C_f=2\tau_w/U_e^2$) illustrated in figure~\ref{fig:BL_props}(f), which is the wall shear stress nondimensionalised by the local reference velocity.

\section{Conditions for self-similarity}
\label{sec:similarity}

To achieve a self-similar boundary layer there are various quantities that must be independent of $x$.
Following the ideas and analysis of \cite{Townsend1956_book,George1993_inbook} and \cite{Castillo2004_JFE}, we start with the Reynolds averaged Navier-Stokes (RANS) continuity, streamwise momentum and wall normal momentum equations equations given by
\begin{eqnarray}
\label{eq:continuity}
  \p_x \langle U \rangle + \p_y \langle V \rangle = 0 \ \mbox{, } \\   
\label{eq:x-mom}
  \langle U \rangle \p_x \langle U \rangle 	
  +\langle V \rangle \p_y \langle U \rangle 
  + \p_x \langle P \rangle 
  + \p_x \langle u u	 \rangle 
  + \p_y \langle u v \rangle 
  - \nu \p_y\p_y \langle U \rangle   
  = \nu \p_x\p_x \langle U \rangle   
  \ \mbox{, } \\
\label{eq:y-mom}
  \p_y \langle P \rangle
  + \p_y \langle v v \rangle 
  = -\langle U \rangle \p_x \langle V \rangle 
  -\langle V \rangle \p_y \langle V \rangle 
  - \p_x \langle u v	 \rangle 
  + \nu \p_x\p_x \langle V \rangle   
  + \nu \p_y\p_y \langle V \rangle   \ \mbox{,}
\end{eqnarray}
\noindent respectively.
Note we have set the density to unity in the above equations and throughout the paper.
In the thin shear layer approximation, the terms on the right of the equals sign in (\ref{eq:x-mom}) and (\ref{eq:y-mom}) are assumed to be negligible \citep{Pope2000_book}.
Integrating the thin shear layer version of (\ref{eq:y-mom}) with respect to $y$, returns $\langle P \rangle = P_e - \langle vv \rangle$, where $P_e$ is the streamwise dependent farfield pressure.
An expression for $\langle V \rangle$ is also attained by integrating (\ref{eq:continuity}) with respect to $y$.
Substituting these results into the thin shear layer version of (\ref{eq:x-mom}), produces the momentum equation
\begin{eqnarray}
\nonumber
\label{eq:momentum}
  \langle U \rangle \p_x \langle U \rangle 
  - \int_0^{y} \p_x \langle U \rangle (x,\tilde{y}) d \tilde{y} \ \p_y \langle U \rangle 
  &=& U_e U_e^\prime \\
  + \p_x \langle v v \rangle 
  &-& \p_x \langle u u	 \rangle 
  - \p_y \langle u v \rangle 
  + \nu \p_y\p_y \langle U \rangle   \ \mbox{.}
\end{eqnarray}
\noindent The $^\prime$ operator represents streamwise derivatives of quantities which are only a function of $x$.

The conditions for self-similarity are determined by expanding the momentum equation (\ref{eq:momentum}), using the following similarity ansatz
\begin{eqnarray}
\label{eq:u_avg_decomp}
  \langle U \rangle(x,y) &=& U_e(x) + U_0(x) \ f(\zeta) \ \mbox{,} \\
\label{eq:uv_decomp}
  \langle uv \rangle (x,y) &=& -R_{uv}(x) \ r_{uv}(\zeta) \ \mbox{,} \\
\label{eq:uu_decomp}
  \langle u u \rangle (x,y) &=& R_{uu}(x) \ r_{uu}(\zeta) \ \mbox{, } \\
\label{eq:vv_decomp}
  \langle v v \rangle (x,y) &=& R_{vv}(x) \ r_{vv}(\zeta) \ \mbox{, } \\
\label{eq:zeta_def}
  \zeta &=& y / L_0(x) \ \mbox{, where} \\
\label{eq:L0_def}
  L_0(x) &\equiv& \delta_1(x) U_e(x) / U_0(x) \ \mbox{,}
\end{eqnarray}
\noindent and $U_0$ is used to nondimensionalise the velocity deficit.
The integrals from $\zeta=0$ to $\zeta=\delta/L_0$ of the similarity functions for the Reynolds stresses $r_{uv}(\zeta)$, $r_{uu}(\zeta)$ and $r_{vv}(\zeta)$ are all defined to be equal to $1$.
This means the functions $R_{uv}(x)$, $R_{uu}(x)$, and $R_{vv}(x)$ can be determined at each $x$ position from the integrals in the $\zeta$ direction of $-\langle uv \rangle (x,y)$, $\langle u u \rangle (x,y)$ and $\langle v v \rangle (x,y)$, respectively.
As first presented in \cite{George1993_inbook}, by substituting equations (\ref{eq:u_avg_decomp}) to (\ref{eq:vv_decomp}) into (\ref{eq:momentum}), one can determine that the following quantities must be independent of $x$ for the flow to be self-similar
\begin{eqnarray}
  C_{uu} &=& R_{uu}/U_e^2 \ \mbox{,} \\
  C_{vv} &=& R_{vv}/U_e^2 \ \mbox{,} \\
  C_{uv} &=& R_{uv}/ \left(U_e^2 \delta_1^\prime \right) \ \mbox{,} \\
\label{eq:c_nu}
  C_\nu &=& \nu / \left( U_e \delta_1 \delta_1^\prime \right) \ \mbox{, and} \\
\label{eq:Lambda}
  \Lambda &=& -\delta_1 U_e^\prime / \left( U_e \delta_1^\prime \right) 
  = \delta_1 P_e^\prime / \left( U_e^2 \delta_1^\prime \right) 
  = \left( U_p / U_e \right)^2 / \delta_1^\prime \ \mbox{, }
\end{eqnarray}
\noindent with $U_e$ also linearly proportional to $U_0$.
The $\Lambda$ parameter, as defined in \cite{Castillo2004_JFE}, quantifies the relationship between the pressure gradient and the outer velocity scale, with $U_p = \sqrt{ P_e^\prime \delta_1 }$ the pressure velocity of \cite{Mellor1966_JFM}.
Streamwise regions of constant $C_{uu}$, $C_{vv}$, $C_{uv}$ and $C_\nu$ for a given TBL, indicates self-similarity of the $\langle u u \rangle$, $\langle v v \rangle$ and $\langle u v \rangle$, and $\nu \p_y \p_y \langle U \rangle$ profiles, respectively.
The magnitude of these coefficients indicates their relative contribution to determining the self-similarity of the system.
If all conditions are met then the TBL is self-similar throughout the entire wall normal domain.
If all but the $C_\nu$ coefficient is streamwise independent then the scaling applies only to the outer flow.

The above generalised theory of boundary layer self-similarity also reproduces the classical results of linearly expanding boundary layers.
As stated in \cite{Skote1998_FTC} the terms $\Lambda \delta_1^\prime/\delta_1 = - U_e^\prime / U_e$, rearranged from (\ref{eq:Lambda}), can be integrated to yield the relationship $U_e \propto \delta_1^{-\Lambda}$.
For the case of a linearly growing boundary layer, where $\delta_1 \propto Kx$, the proportionality of the reference velocity becomes $U_e \propto (Kx)^{-\Lambda} \propto x^{-\Lambda} \equiv x^{m}$, with $m=-\Lambda$.
Additionally for the incipient separation case, substituting in the empirical values of $U_p/U_e=1/10.27$ and $\delta_1^\prime = K=0.041$ from \cite{Mellor1966_JFM} into (\ref{eq:Lambda}), one can determine the power exponent for linearly growing boundary layers to be $m = -\Lambda \equiv -(U_p/U_e)^2 / \delta_1^\prime = -1/10.27^2/0.041 = -0.23$, which is the expected value for the incipient APG TBL.
Again linking back to the classical theory of \cite{Mellor1966_JFM}, the widely quoted $\beta$ parameter is equivalent to $U_p^2/U_\tau^2$, where for a unit density the friction velocity $U_\tau\equiv\sqrt{\tau_w}$. 
The $\beta$ parameter, however, becomes undefined as one approaches the incipient separation case as $\tau_w$ (and hence $U_\tau$) approaches zero.
For this reason, the self-similarity of the pressure gradient term for all pressure gradients (zero to incipient) is more appropriately assessed via the streamwise dependence of $\Lambda\equiv U_p^2/U_e^2/\delta_1^\prime$, or equivalently $U_p/U_e$ in the case of a linearly growing displacement thickness of constant $\delta_1^\prime$.

\begin{figure}
\centering
\psfrag{ZPG}[c][][1.0]{$\beta=0$}
\psfrag{beta1}[c][][1.0]{$\beta=1$}
\psfrag{betaInf}[c][][1.0]{$\beta=39$}
\psfrag{Cnu*10}[c][b][1.0]{$C_{\nu}(x)\times10$}
\psfrag{Cnu}[c][b][1.0]{$C_{\nu}(x)$}
\psfrag{Cuu*1e2}[c][b][1.0]{$C_{uu}(x)\times10^2$}
\psfrag{Cuv}[c][b][1.0]{$C_{uv}(x)$}
\psfrag{Cvv*1e2}[c][b][1.0]{$C_{vv}(x)\times10^2$}
\psfrag{uP/uRef}[c][][1.0]{$U_P(x)/U_e(x)$}
\psfrag{x/delta(ReDelta=5000)}[c][][1.0]{$(x-x_{\star})/\delta_1(x_{\star})$}
\begin{tabular}{ll}
    (a) & (b) \\
    \includegraphics[trim = 3mm 0mm 3mm 2mm, clip, width=0.5\textwidth]{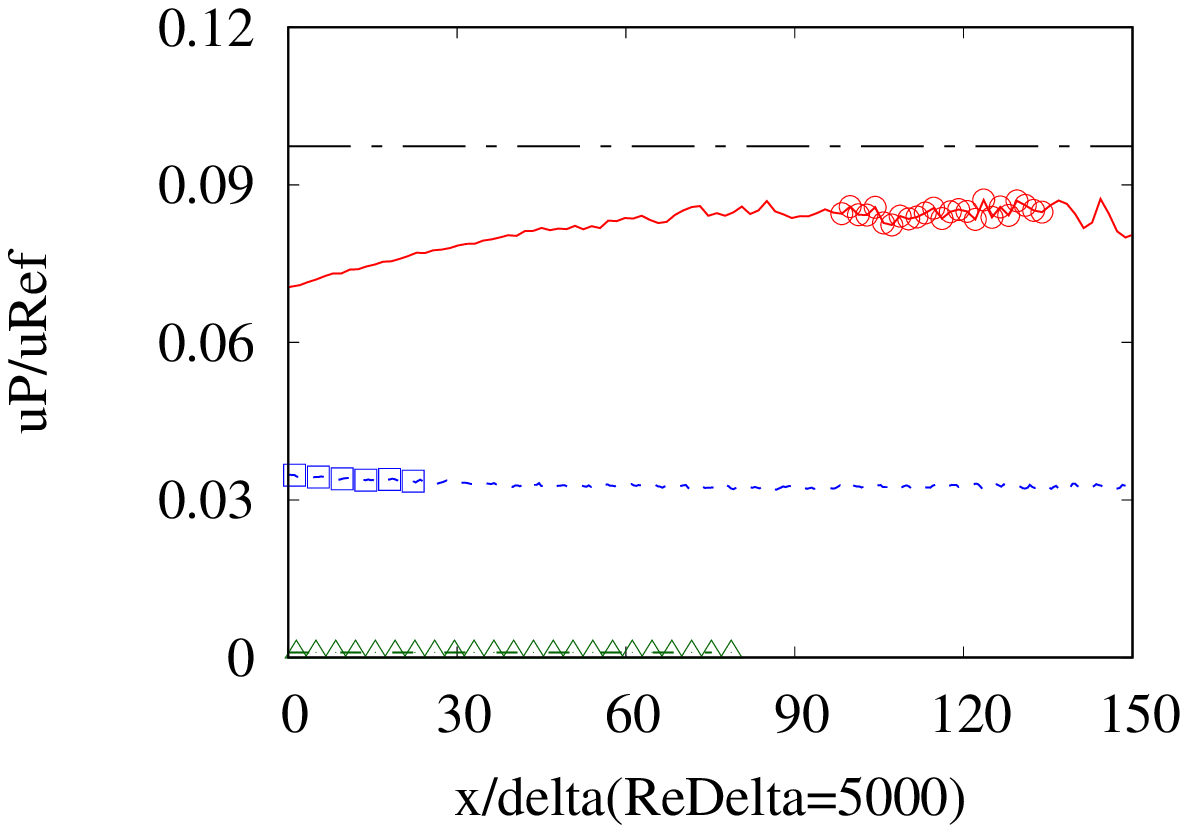} &
   \includegraphics[trim = 3mm 0mm 3mm 2mm, clip, width=0.5\textwidth]{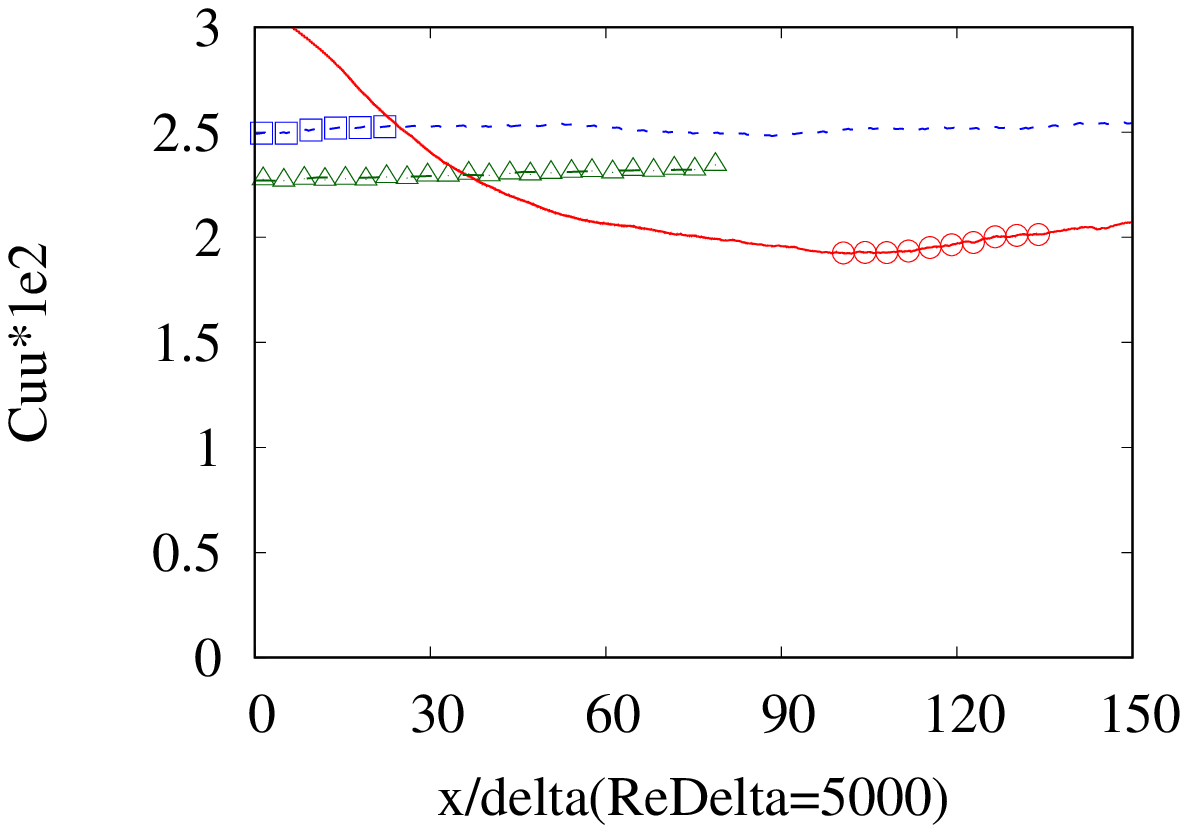} \\
    (c) & (d) \\
    \includegraphics[trim = 3mm 0mm 3mm 2mm, clip, width=0.5\textwidth]{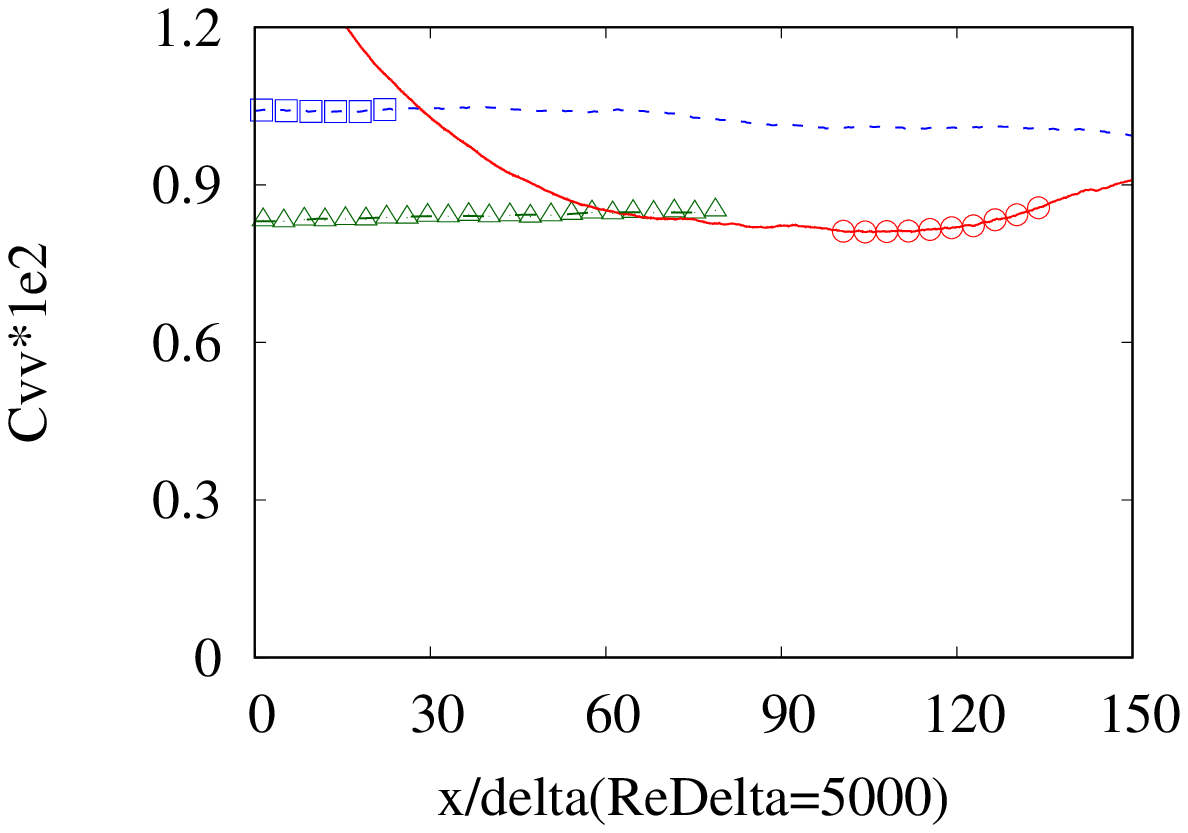} &
    \includegraphics[trim = 3mm 0mm 3mm 2mm, clip, width=0.5\textwidth]{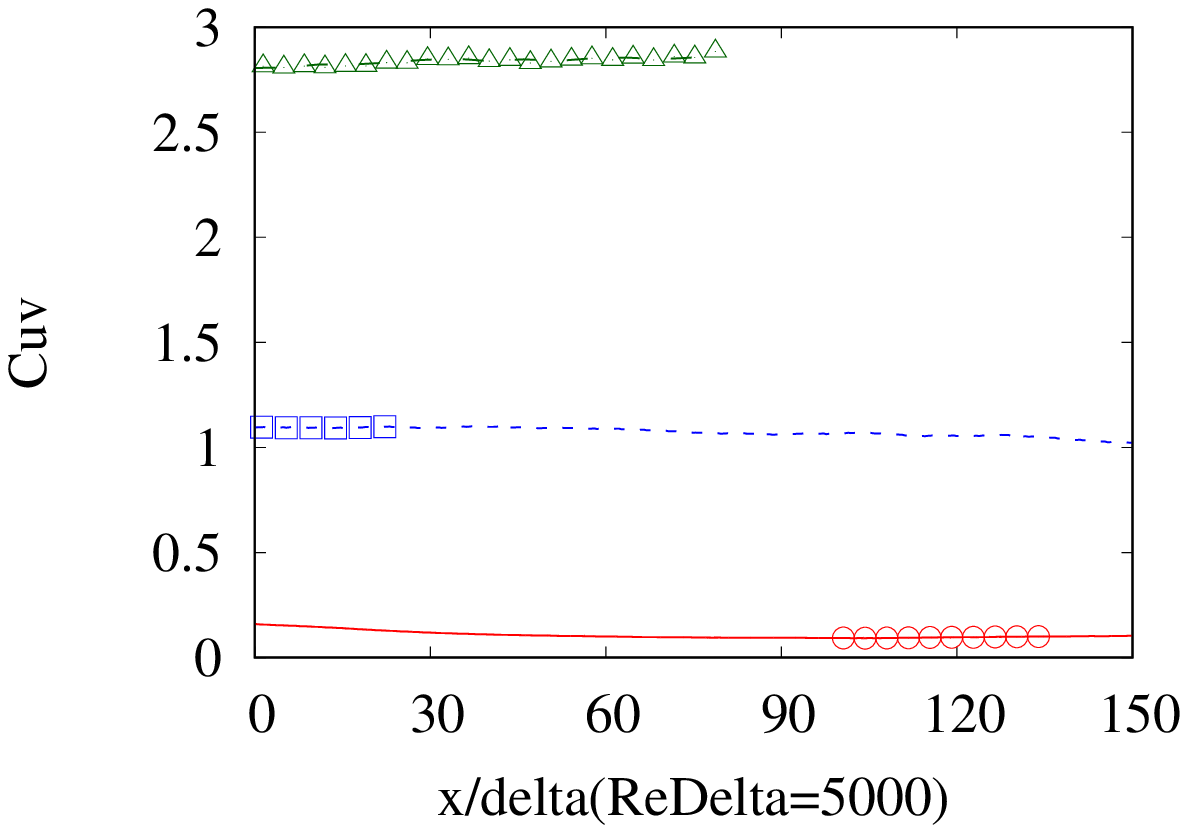} \\
    \includegraphics[trim = 3mm 0mm 3mm 2mm, clip, width=0.5\textwidth]{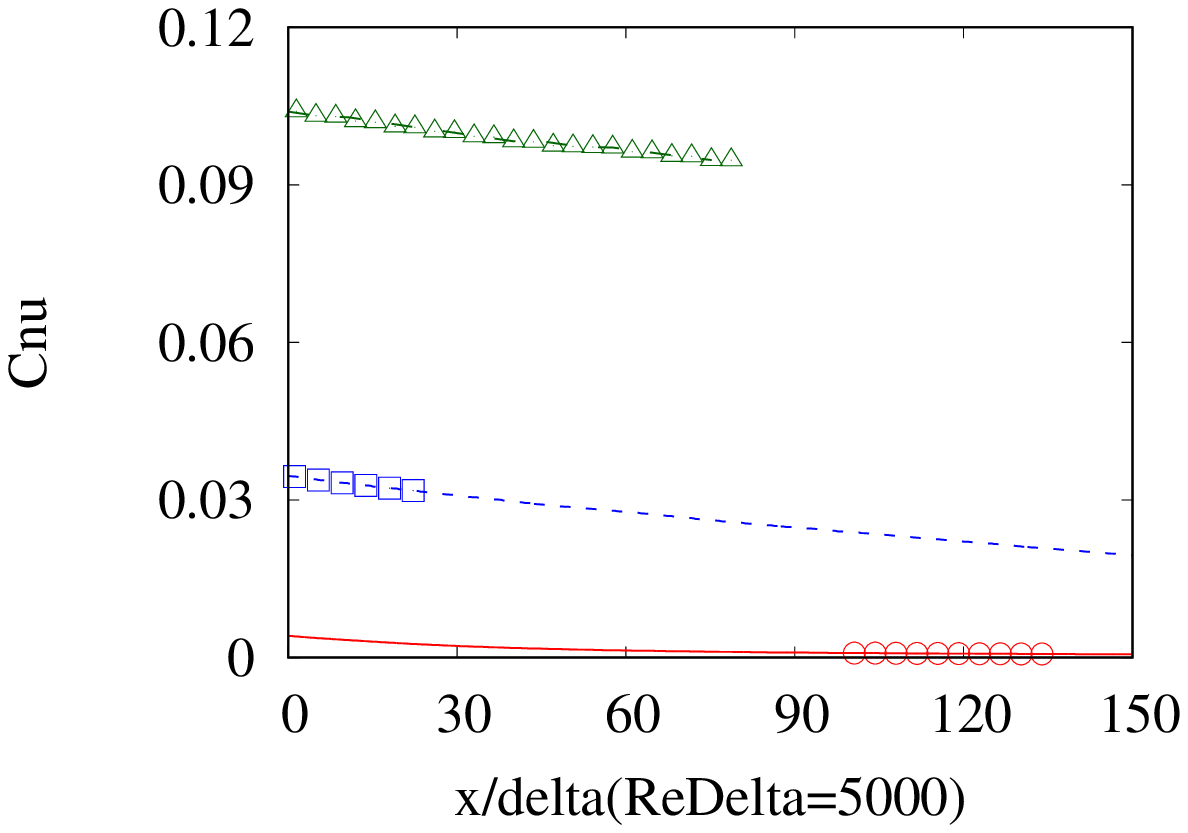} \\
\end{tabular}
\caption{Coefficients to assess the self-similarity of the strong APG (solid red lines, with domain of interest indicated by \textcolor{red}{$\circ$}), 
mild APG (short dashed blue lines, with domain of interest indicated by \textcolor{blue}{\opensquare}) and 
ZPG (long dashed green lines, with domain of interest indicated by \textcolor{green}{\opentriangle}) TBL on the basis of:
(a) pressure gradient velocity ratio, $U_P/U_e$, with empirical values of $U_P/U_e=0.097$ \citep{Mellor1966_JFM} indicated by the black dash-dotted line;
(b) $C_{uu}$;
(c) $C_{vv}$; 
(d) $C_{uv}$; and
(e) $C_{\nu}$.
Note, $x_{\star}$ is the streamwise position at which $Re_{\delta_1}=4800$.
}
\label{fig:coeff}
\end{figure}

Each of the similarity coefficients are now discussed in terms of both their relative magnitude and streamwise independence.
For each TBL the coefficients $U_P/U_e$, $C_{uu}$, $C_{vv}$, $C_{uv}$, and $C_\nu$ are illustrated in figure~\ref{fig:coeff}(a)-(e), respectively.
Their streamwise averaged values and standard deviations over the domain of interest are also listed in table~\ref{table:similarity}.
The magnitude of the similarity coefficients indicate the importance of each of the terms in the boundary layer equations.
As one would expect $U_P/U_e$ increases with pressure gradient.
For the strong APG TBL $U_P/U_e$ is also within $9\%$ of the empirical value of the incipient case of $1/10.27=0.097$ from \cite{Mellor1966_JFM}.
The coefficients $C_{uu}$ and $C_{vv}$ are of the same order of magnitude for all TBL.
The term $C_{uv}$ decreases with pressure gradient due to the increased slope in the displacement thickness.
Note for the calculation of $C_{uv}$, the slope of the displacment thickness is quite noisy, so we use a value streamwise averaged over the respective domains of interest.
For each TBL the coefficients $U_P/U_e$, $C_{uu}$, $C_{vv}$, and $C_{uv}$, are all relatively constant over the domain of interest, with standard deviations at worst $2.5$\% of their associated streamwise averaged value (as derived from table~\ref{table:similarity}).
The only term that is a strong function of $x$ is $C_\nu$, see figure~\ref{fig:coeff}(e), which measures the self-similarity of the viscous term in the boundary layer streamwise momentum equation (\ref{eq:momentum}).
However, as can be seen in table~\ref{table:similarity}, the streamwise average value of $C_{\nu}$ decreases monotonically as the pressure gradient increases.
The magnitude of $C_{\nu}$ relative to each of the other coefficients (eg: $C_\nu/C_{uu}$) also decreases as the pressure gradient increases, indicating that the viscous term is becoming a weaker constraint on self-similarity.
In fact, the mean and Reynolds stress profiles for the strong APG TBL are shown in section~\ref{sec:mean_profiles} and section~\ref{sec:Reynolds_stresses} to collapse at different streamwise positions.

\begin{table}
\caption{Streamwise average of the similarity variables within the domain of interest for each TBL.
The quantity in each parentheses is the associated streamwise standard deviation.}
\centering
\begin{tabular}{cccc}
  & & & \\
  \hline
  & $ \ \ \ \ \ \ \ \ $ ZPG $ \ \ \ \ \ \ \ \ $ & $ \ \ \ \ \ $ Mild APG $ \ \ \ \ \ $ & $ \ \ \ $ Strong APG $ \ \ \ $ \\
  \hline
  nominal $\beta$ & $0$ & $1$ & $39$ \\ 
  $U_P/U_e$ 	& $0 	\ (0)$ 				  & $3.3\times10^{-2} \ (2.6\times10^{-4})$ & $8.5\times10^{-2} \ (1.3\times10^{-3})$ \\    
  $C_{uu}$ 	& $2.3\times10^{-2} \ (1.4\times10^{-4})$ & $2.5\times10^{-2} \ (2.8\times10^{-5})$ & $1.9\times10^{-2} \ (3.2\times10^{-4})$ \\
  $C_{vv}$ 	& $8.4\times10^{-3} \ (4.4\times10^{-5})$ & $1.0\times10^{-2} \ (2.2\times10^{-5})$ & $8.2\times10^{-3} \ (1.2\times10^{-4})$ \\
  $C_{uv}$ 	& $2.8 \ (1.3\times10^{-2})$ 		  & $1.1 	      \ (2.5\times10^{-3})$ & $8.9\times10^{-2} \ (2.2\times10^{-3})$ \\
  $C_{\nu}$ 	& $9.9\times10^{-2} \ (2.1\times10^{-3})$ & $3.0\times10^{-2} \ (7.8\times10^{-4})$ & $7.3\times10^{-4} \ (5.6\times10^{-5})$ \\  
  \hline
\end{tabular}
\label{table:similarity}
\end{table}

\section{Mean streamwise velocity}
\label{sec:mean_profiles}

In this section the self-similarity of the strong APG TBL is first assessed based upon the collapse of the mean streamwise velocity profiles.
This is followed by the impact that the pressure gradient has on the log-layer, viscous sub-layer and mean field inflection points.

\begin{figure*}
\psfrag{ZPG}[c][][1.0]{$\beta=0$}
\psfrag{beta1}[c][][1.0]{$\beta=1$}
\psfrag{betaInf}[c][][1.0]{$\beta=39$}
\psfrag{y/delta1}[c][][1.0]{$y / \delta_1$}
\psfrag{dUdy/Uref*delta1}[c][b][1.0]{$\p_y \langle U \rangle \ \delta_1 / U_e$}
\psfrag{U/Uref}[c][b][1.0]{$\langle U \rangle/U_e$}
\begin{center}
  \begin{tabular}{ll}
  (a) & (b) \\
 \includegraphics[trim = 3mm 0mm 3mm 2mm, clip, width=0.5\textwidth]{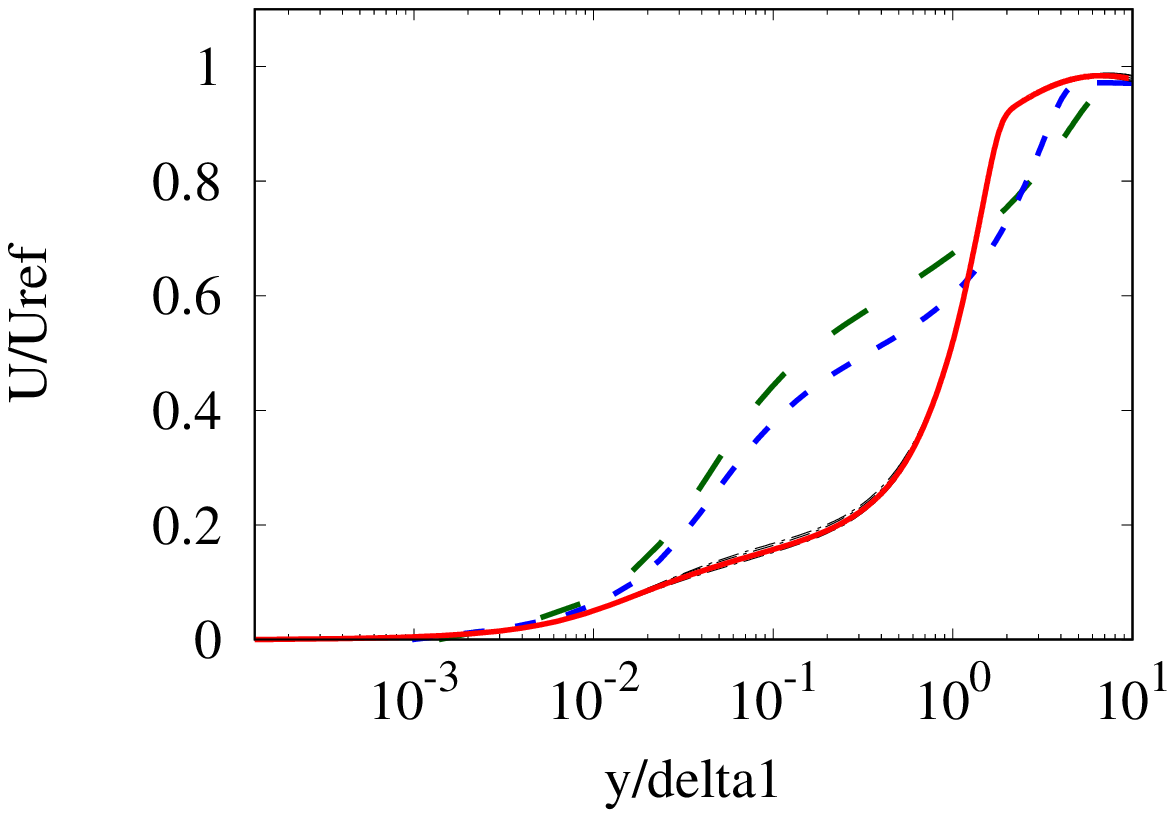} &
 \includegraphics[trim = 3mm 0mm 3mm 2mm, clip, width=0.5\textwidth]{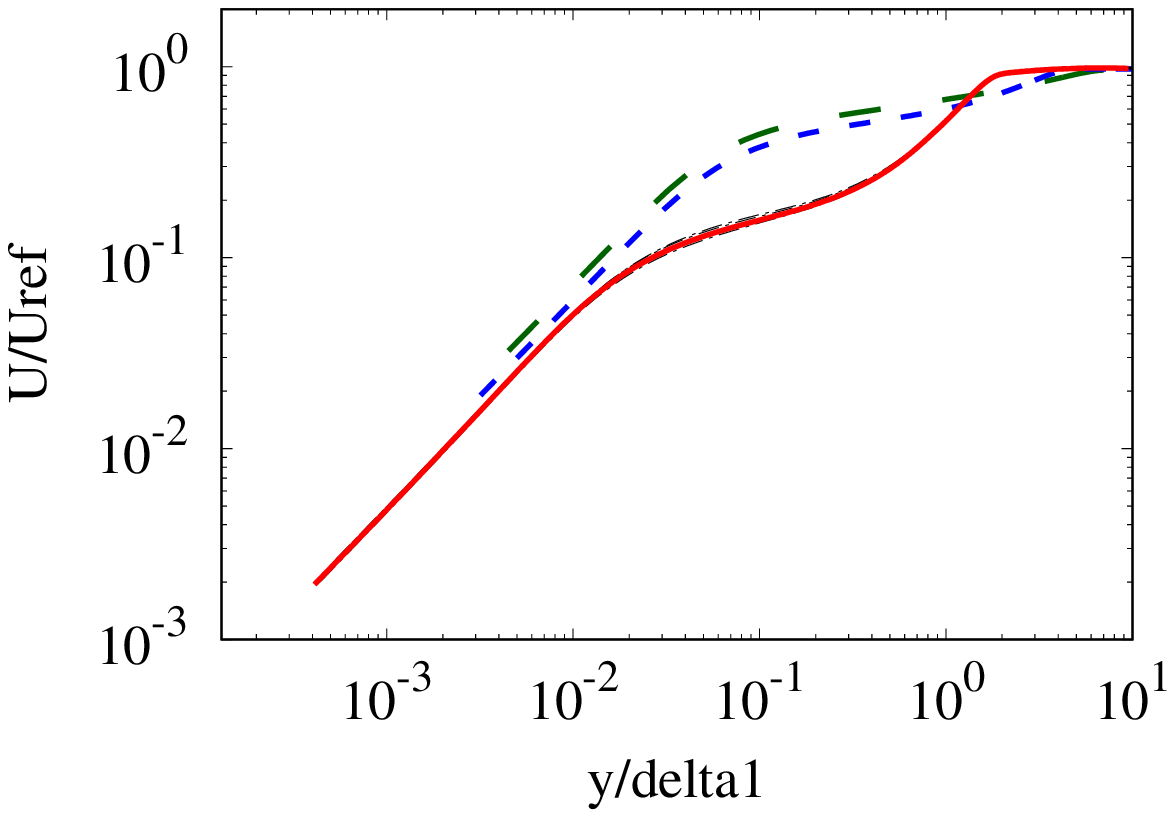} \\
  (c) & \\
 \includegraphics[trim = 3mm 0mm 3mm 2mm, clip, width=0.5\textwidth]{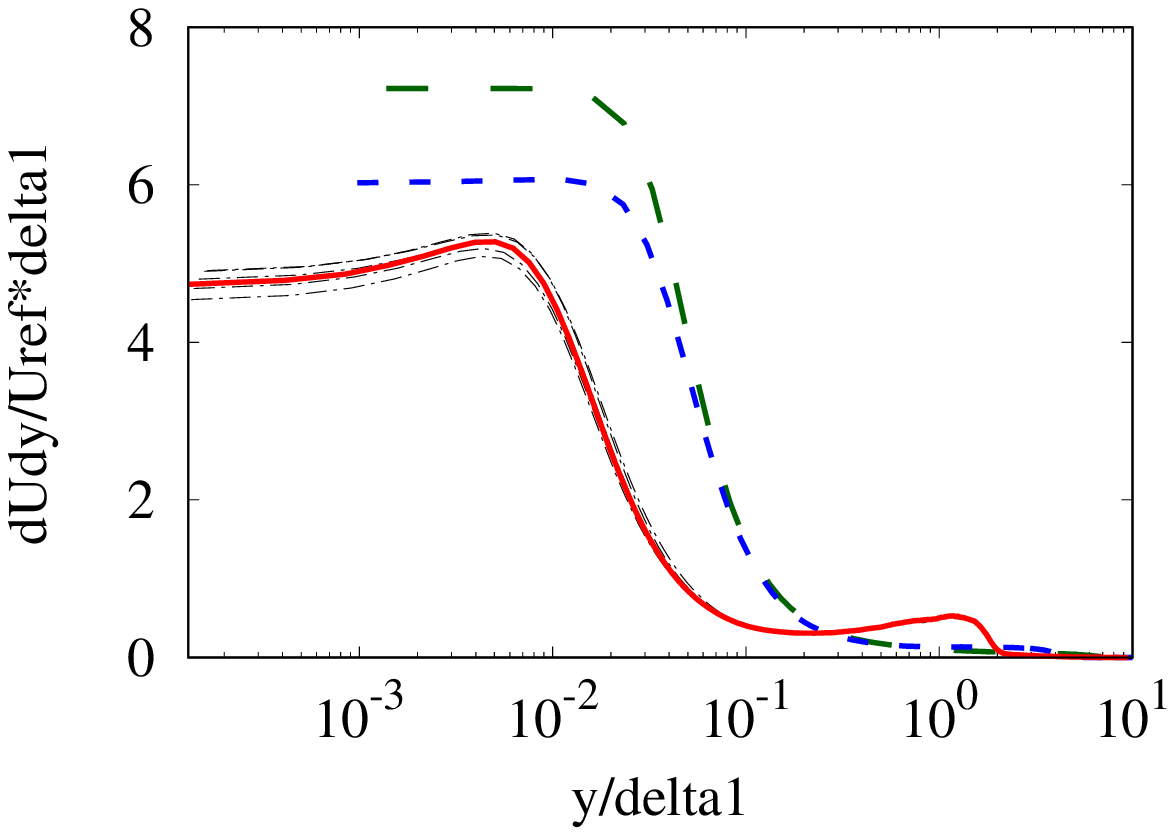} & \\
  \end{tabular}
\end{center}
\caption{Streamwise averaged statistical profiles nondimensionalised by $U_e$ and $\delta_1$ of the strong APG (solid red lines), mild APG (short dashed blue lines) and ZPG (long dashed green lines) TBL DNS:
(a) log-linear plot of $\langle U \rangle$;
(b) log-log plot of $\langle U \rangle$; and
(c) log-linear plot of $\p_y \langle U \rangle$.
The dash-dotted black lines are profiles of the strong APG at individual streamwise positions.
}
\label{fig:mean}
\end{figure*}

The self-similarity is assessed by comparing the mean streamwise velocity profiles ($\langle U \rangle$) at various streamwise stations.
The velocity profiles are nondimensionalised by the local $U_e$, and the wall normal position by the local $\delta_1$.
The dash-dotted black lines in figure~\ref{fig:mean}(a) are profiles from the strong APG TBL at equally space streamwise locations within the domain of interest.
These profiles collapse under this scaling, indicating the mean field is self-similar.
The solid red line represents the streamwise average in scaled coordinates throughout this streamwise domain.

The impact of the pressure gradient on the log-layer is illustrated by comparing the $\langle U \rangle$ profiles across each of the TBL.
The streamwise averaged profiles within their respective domains of interest are illustrated in figure~\ref{fig:mean}(a) for the ZPG (long dashed green line), mild APG (short dashed blue line), and the strong APG TBL (solid red line).
From a comparison of these profiles, it is clear that as the pressure gradient increases, the wall normal extent of the wake region increases and the extent of the log-layer decreases.
The existence of a log-layer is derived on the basis that there is a wall normal region within which there are two important length scales: an inner scale based on the wall shear stress; and an outer scale based on a measure of the boundary layer thickness ($\delta_1$ for example).
The fact that the log-layer is almost non-existent for the strong APG case is indicative that for the vast majority of the wall normal domain, there is only one pertinent length scale, the outer scale.
Note that a theoretical velocity profile valid in the log-layer was derived in \cite{Skote2002_JFM} for both attached and separated boundary layer flows, which compared well with their DNS data as well as with more recent simulations \citep{Cheng2015_JFM}.

The viscous sub-layer is also shown to reduce in extent as the pressure gradient increases.
Figure~\ref{fig:mean}(b) presents the same profiles as those discussed above but with log-log axes.
For all of the TBL, there is a linear relationship between $y$ and $\langle U \rangle$ in the near wall region.
A linear relationship implies that $\nu/U_\tau$ is the only important length scale in this zone.
For the ZPG and mild APG case, this linear region appears to end at $y\approx 0.05\delta_1$.
The linear region in the strong APG case ends an order of magnitude closer to the wall.
These observations can be explained by deriving the functional form of $\langle U \rangle$ from the equations of motion, specific to the viscous sub-layer.
Let us make the standard assumptions that within the viscous sub-layer the Reynolds stress terms and advective terms are also negligible in the boundary form of the momentum equations, which reduces (\ref{eq:x-mom}) to $P_e^\prime = \nu\p_y\p_y\langle U \rangle$.
Integrating this expression with respect to $y$ twice, and applying the boundary conditions $\langle U \rangle (y=0)=0$ and $\mu \p_y \langle U \rangle (y=0)=\tau_w = U_\tau^2$ returns
\begin{eqnarray}
\label{eq:sublayer}
  \langle U \rangle (y) = \f{1}{2\nu}P_e^\prime y^2 + \f{1}{\nu} U_\tau^2 y \ \mbox{,}
\end{eqnarray}
\noindent where $\mu$ is the dynamic viscosity.
Substituting the definition of the pressure velocity, $U_p = \sqrt{ P_e^\prime \delta_1 }$, into (\ref{eq:sublayer}) then gives
\begin{eqnarray}
\label{eq:sublayer_nondim}
  \langle U \rangle (y) = \f{\delta_1}{\nu} \left[ \f{1}{2}U_P^2 \left(\f{y}{\delta_1}\right)^2 + U_\tau^2 \f{y}{\delta_1} \right] \ \mbox{.}
\end{eqnarray}
\noindent One could equally chose to define pressure gradient based length and velocity scales as per \cite{Skote2002_JFM}, however, we make the above choice for consistency with the analysis in section~\ref{sec:similarity}.
In the limit of zero pressure gradient, $\langle U \rangle / U_\tau =  y U_\tau/\nu$ (or equivalently $U^+ = y^+$), with $\langle U \rangle$ proportional to $y$.
In the limit of zero mean wall shear stress (incipient separation), $\langle U \rangle / U_P =  y^2 U_P/(2\nu\delta_1)$, with $\langle U \rangle$ proportional to $y^2$.
As the pressure gradient increases with respect to the wall shear stress, within the viscous sub-layer $\langle U \rangle$ transitions from being a linear function of $y$ to a quadratic one.

Finally the impact of the pressure gradient of the location of the inflection points is illustrated by the wall normal gradient of the mean streamwise velocity profiles.
The wall normal gradient of $\langle U \rangle$ is illustrated in figure~\ref{fig:mean}(c) again scaled in terms of the local $\delta_1$ and $U_e$.
There are two evident points of inflection in the APG TBL, one in the near wall region, and another at the approximate height of the displacement thickness.
These points of inflection coincide with the inner and outer peaks of turbulent production, which is discussed further in section~\ref{sec:budget_profiles}.
As can be seen the gradient at the wall for the strong APG case is smaller than in the other TBL in nondimensional terms, but not zero.
This is again consistent with the model of the viscous sub-layer in (\ref{eq:sublayer_nondim}).
The relative contribution of the pressure gradient to the wall shear stress term is given by the ratio
\begin{eqnarray}
\label{eq:sublayer_nondim_ratio}
  \f{U_P^2 \ (y/\delta_1)^2 / 2}{U_\tau^2 \ (y/\delta_1)}  =  \f{U_P^2}{U_\tau^2} \f{y}{2\delta_1}  = \beta \f{y}{2\delta_1} \ \mbox{. }
\end{eqnarray}
\noindent As discussed above, (\ref{eq:sublayer_nondim_ratio}) indicates that as the pressure gradient relative to the shear stress increases (quantified by $\beta$), the contribution of the $U_P$ term increases.
Importantly this expression also indicates that as one approaches the wall, the contribution of the pressure gradient term decreases.
In fact for $y/\delta_1 \gg 2/\beta$ the pressure gradient dominates, and for $y/\delta_1 \ll 2/\beta$ the viscous terms dominates.
Therefore, in all but the limiting incipient separation case of infinite $\beta$, there will always be a region in which $\langle U \rangle$ is linearly related to $y$.

\section{Reynolds stresses}
\label{sec:Reynolds_stresses}

The self-similarity of the strong APG TBL is now assessed based upon the collapse of the Reynolds stress profiles.
The impact that the pressure gradient has on the location and magnitude of the inner and outer peaks is then discussed, followed by a proposed physical explanation based upon linear stability arguments.

\begin{figure*}
\psfrag{ZPG}[c][][1.0]{$\beta=0$}
\psfrag{beta1}[c][][1.0]{$\beta=1$}
\psfrag{betaInf}[c][][1.0]{$\beta=39$}
\psfrag{<u'u'>/Uref/Uref*1e2}[c][b][1.0]{$\langle u u \rangle/U_e^2\times10^2$}
\psfrag{<v'v'>/Uref/Uref*1e2}[c][b][1.0]{$\langle v v \rangle/U_e^2\times10^2$}
\psfrag{<w'w'>/Uref/Uref*1e2}[c][b][1.0]{$\langle w w \rangle/U_e^2\times10^2$}
\psfrag{-<u'v'>/Uref/Uref/dD1dx}[c][b][1.0]{$-\langle u v \rangle/\left( U_e^2 \ \delta_1^\prime \right)$}
\psfrag{<u'v'>/Uref/Uref}[c][b][1.0]{$-\langle u v \rangle/U_e^2\times10^2$}
\psfrag{y/delta1}[c][][1.0]{$y / \delta_1$}
\begin{center}
  \begin{tabular}{ll}
  (a) & (b) \\
 \includegraphics[trim = 3mm 0mm 3mm 2mm, clip, width=0.5\textwidth]{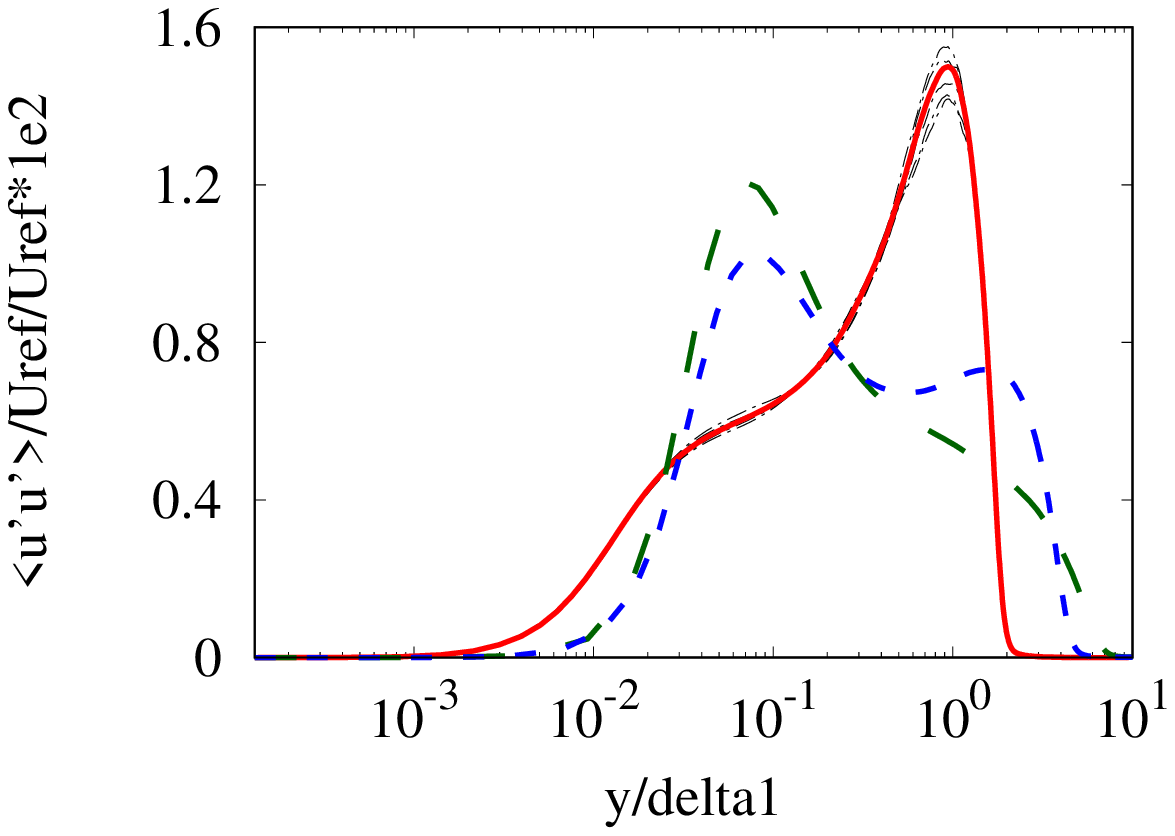} &
 \includegraphics[trim = 3mm 0mm 3mm 2mm, clip, width=0.5\textwidth]{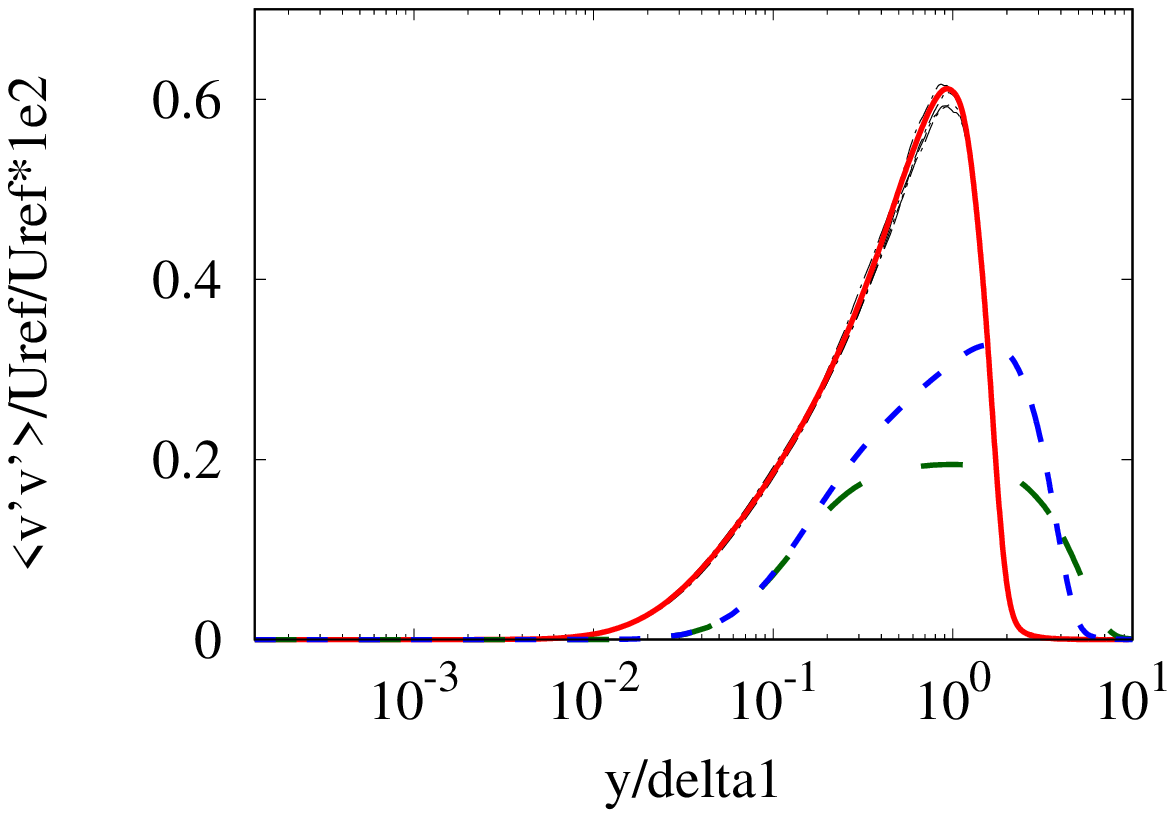} \\
  (c) & (d) \\
 \includegraphics[trim = 3mm 0mm 3mm 2mm, clip, width=0.5\textwidth]{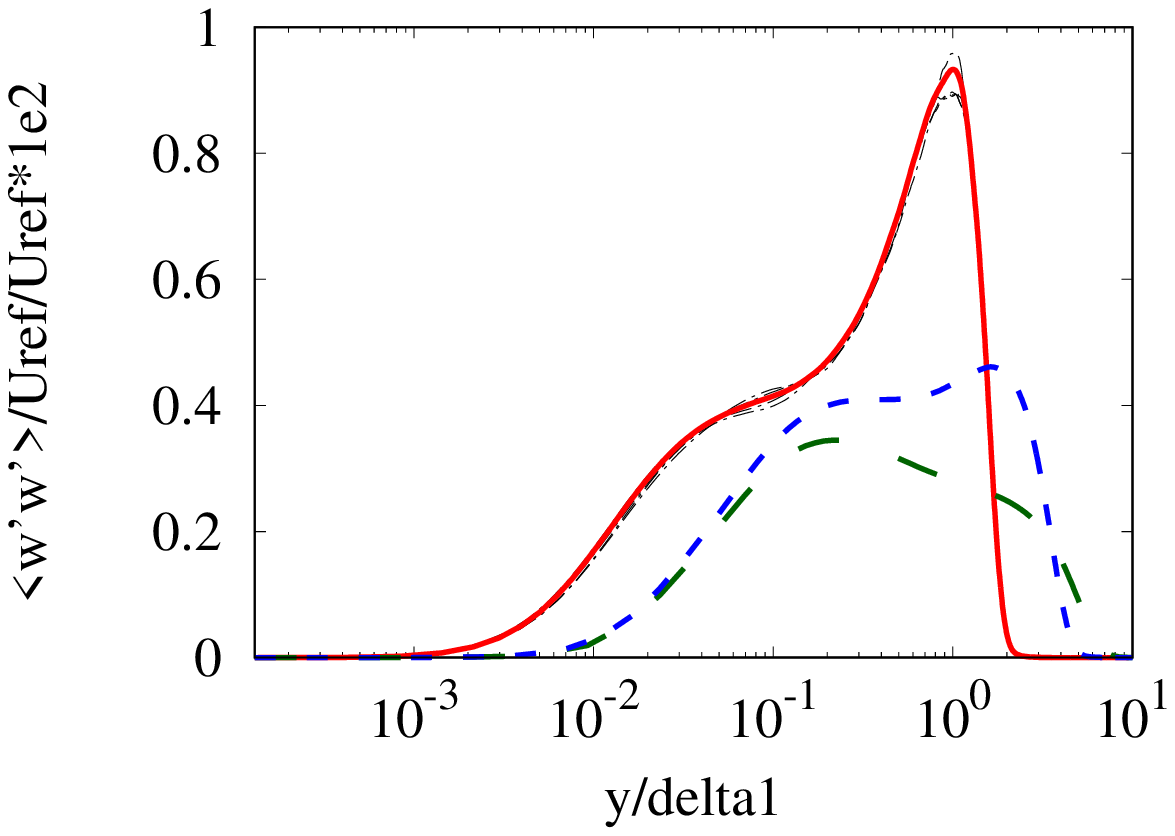} & 
 \includegraphics[trim = 3mm 0mm 3mm 2mm, clip, width=0.5\textwidth]{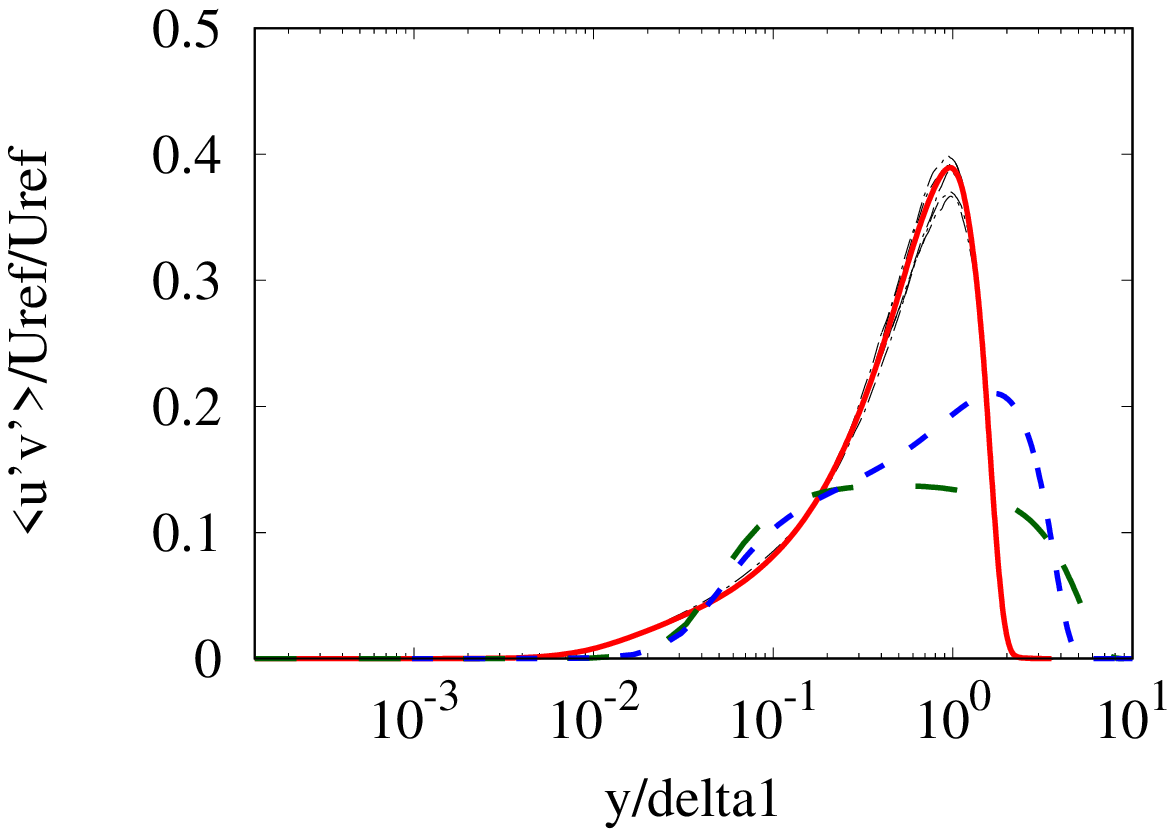} \\ 
  \end{tabular}
\end{center}
\caption{Streamwise averaged statistical profiles nondimensionalised by $U_e$ and $\delta_1$ of the strong APG (solid red lines), mild APG (short dashed blue lines) and ZPG (long dashed green lines) TBL DNS:
(a) $\langle u u \rangle$ ;
(b) $\langle v v \rangle$ ; 
(c) $\langle w w \rangle$ ; and 
(d) $-\langle u v \rangle$.
The dash-dotted black lines are profiles of the strong APG at individual streamwise positions.
}
\label{fig:BL_stresses}
\end{figure*}

As was undertaken for the mean velocity profiles, the Reynolds stresses are presented scaled on the basis of $U_e$ and $\delta_1$.
Profiles of $\langle u u \rangle$, $\langle v v \rangle$, $\langle w w \rangle$, and $\langle u v \rangle$, are respectively illustrated in figures~\ref{fig:BL_stresses}(a)-(d), for the ZPG, mild APG and strong APG TBL.
Note, according to the theoretical framework presented in section~\ref{sec:similarity}, the profiles of $\langle u v \rangle$ in figure~\ref{fig:BL_stresses}(d) should in fact be nondimensionalised by $U_e^2 \delta_1^\prime$ instead of $U_e^2$.
However, we adopt the latter scaling for consistency with the other Reynolds stresses, in order to clearly illustrate their relative magnitudes.
Each of the individual profiles of the strong APG case (dash-dotted black lines) collapse within the domain of interest over most of the wall normal domain.
As expected the largest spread is located at the point of maximum variance (ie. the outer peak).
Additional temporal sampling would reduce the variation across the profiles.

The existence, location and magnitude of the outer peaks are strongly dependent upon the pressure gradient, whilst the properties of the inner peak are Reynolds number dependent.
The inner peak of $\langle u u \rangle$ and $\langle w w \rangle$ for the ZPG and mild APG cases are located at similar distances from the wall, due to the streamwise averaging undertaken over the same $Re_{\delta_1}$ range.
No inner peak is evident in the profiles of $\langle v v \rangle$ and $\langle u v \rangle$, as it is dominated by the presence of the outer peak.
An outer peak is evident in all of the Reynolds stresses for the mild and strong APG TBL, and becomes more evident as the pressure gradient increases.
The outer peak in all of the Reynolds stresses is located at $y=\delta_1$ for the strong APG TBL, and at $y=1.3\delta_1$ for the mild APG flow.
For both TBLs it is the same position as their respective outer inflection points in their mean streamwise velocity profiles illustrated in figure~\ref{fig:mean}(c).
The coincidence of the outer peak in the Reynolds stresses and inflection point in the mean velocity profile, has also been observed in previous APG TBL DNS \citep{Araya2013_PF}.
This observation suggests that a shear flow instability is the dominant mechanism contributing to the fluctuations in the outer region.

Linear stability theory provides an explanation for the location of the inner and outer peaks in the Reynolds stresses.
Take the two limiting cases of the incipient APG TBL and the ZPG TBL.
The incipient APG TBL has zero mean shear stress at the wall, with the only source of mean shear being that imparted by the pressure gradient.
The stability properties in this case are analogous to those of a free shear layer, which would generate fluctuations in all velocity components across all spanwise and streamwise scales distributed about the point of inflection.
For the ZPG TBL the only source of shear is the wall itself.
Linear stability theory has also identified modes that represent the fluctuations of the near wall and outer regions in analogous turbulent channel flows \citep{Alamo2006_JFM, Kitsios2010_JFM}.
Fluctuations generated by such instabilities imprint themselves as peaks in the Reynolds stress profiles.
Arguably as the pressure gradient increases the flow starts to behave less like a ZPG TBL and more like a free shear layer.
The Reynolds stresses centred at the outer point of inflection (outer peak), would then begin to dominate over any wall driven shear instabilities (inner peak), which is precisely what is observed.
This is a somewhat simplified view, since the linear stability properties of two separate flows with different background states (i.e. ZPG TBL and free shear layer) cannot simply be superimposed.
The stability properties are dependent on the details of the base flow.
The transfer of momentum between these turbulent fluctuations and the mean field is discussed in section~\ref{sec:momentum_terms}.
These fluctuations are also transferred throughout the wall normal domain via nonlinear process quantified by the turbulent kinetic energy transfer term, as presented in section~\ref{sec:budget_profiles}.

\section{Streamwise and wall normal momentum terms}
\label{sec:momentum_terms}

\begin{figure}
\psfrag{y/delta1}[c][b][1.0]{$y/\delta_1$}
  \psfrag{u vX}[c][][1.0]{\scriptsize $\langle U \rangle \p_x \langle V \rangle \ \ \ \ $}
  \psfrag{v vY}[c][][1.0]{\scriptsize $\langle V \rangle \p_y \langle V \rangle \ \ \ \ $}
  \psfrag{pY}[c][][1.0]{\scriptsize $\p_y \langle P \rangle \ \ \ \ $}
  \psfrag{uvX}[c][][1.0]{\scriptsize $\p_x \langle uv \rangle \ \ \ \ $}
  \psfrag{vvY}[c][][1.0]{\scriptsize $\p_y \langle vv \rangle \ \ \ \ $}
  \psfrag{-nu vXX}[c][][1.0]{\scriptsize $-\nu \p_x \p_x \langle V \rangle \ \ \ \ $}
  \psfrag{-nu vYY}[c][][1.0]{\scriptsize $-\nu \p_y \p_y \langle V \rangle \ \ \ \ $}
  \psfrag{u uX}[c][][1.0]{\scriptsize $\langle U \rangle \p_x \langle U \rangle \ \ \ \ $}
  \psfrag{v uY}[c][][1.0]{\scriptsize $\langle V \rangle \p_y \langle U \rangle \ \ \ \ $}
  \psfrag{pX}[c][][1.0]{\scriptsize $\p_x \langle P \rangle \ \ \ \ $}
  \psfrag{uvY}[c][][1.0]{\scriptsize $\p_y \langle uv \rangle \ \ \ \ $}
  \psfrag{uuX}[c][][1.0]{\scriptsize $\p_x \langle uu \rangle \ \ \ \ $}
  \psfrag{-nu uXX}[c][][1.0]{\scriptsize $-\nu \p_x \p_x \langle U \rangle \ \ \ \ $}
  \psfrag{-nu uYY}[c][][1.0]{\scriptsize $-\nu \p_y \p_y \langle U \rangle \ \ \ \ $}
  \psfrag{res}[c][][1.0]{\scriptsize residual $ \ \ \ \ \ \ $}
\begin{center}
  \begin{tabular}{ll}
  (a) & (b) \\
  \psfrag{momX}[c][b][1.0]{$x$-momentum terms $\times10^2$}
  \includegraphics[trim = 11mm 0mm 6mm 0mm, clip, height=0.41\textwidth]{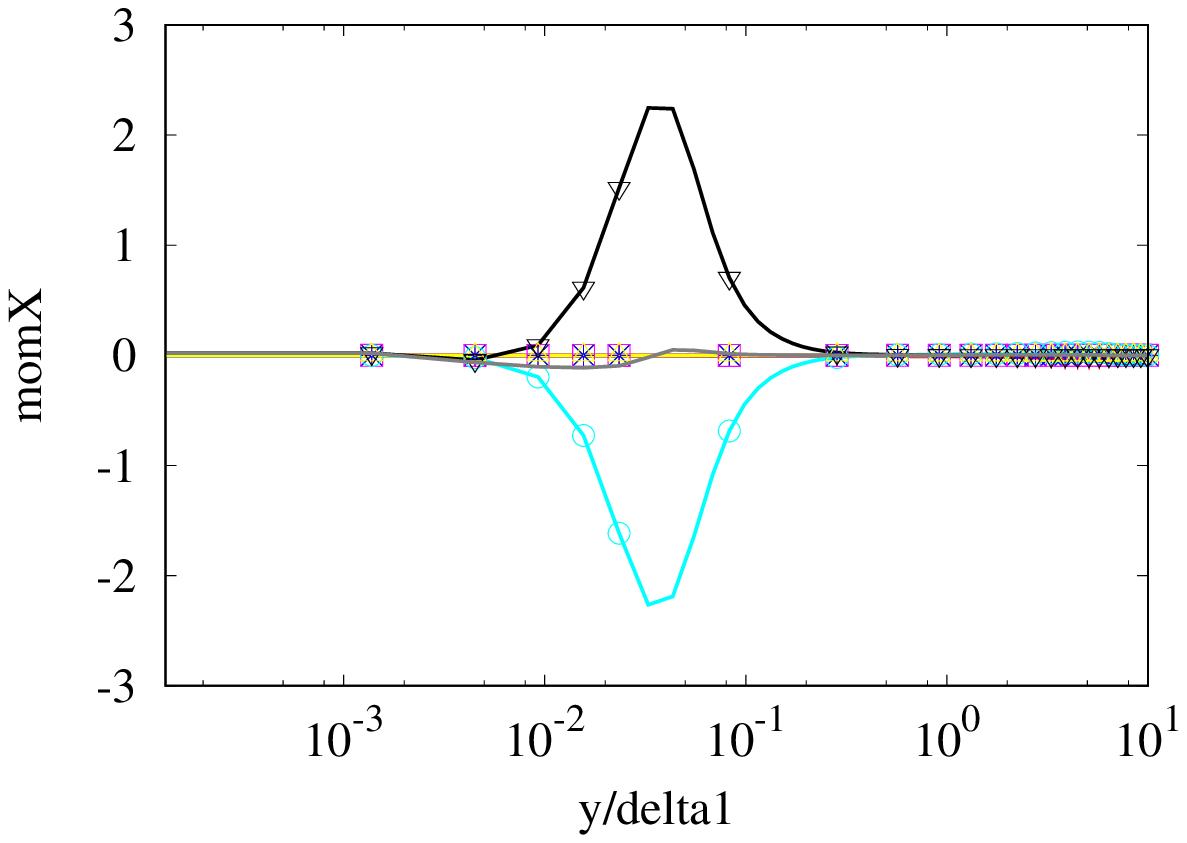} &
  \psfrag{momX}[c][b][1.0]{}
  \includegraphics[trim = 18mm 0mm 6mm 0mm, clip, height=0.41\textwidth]{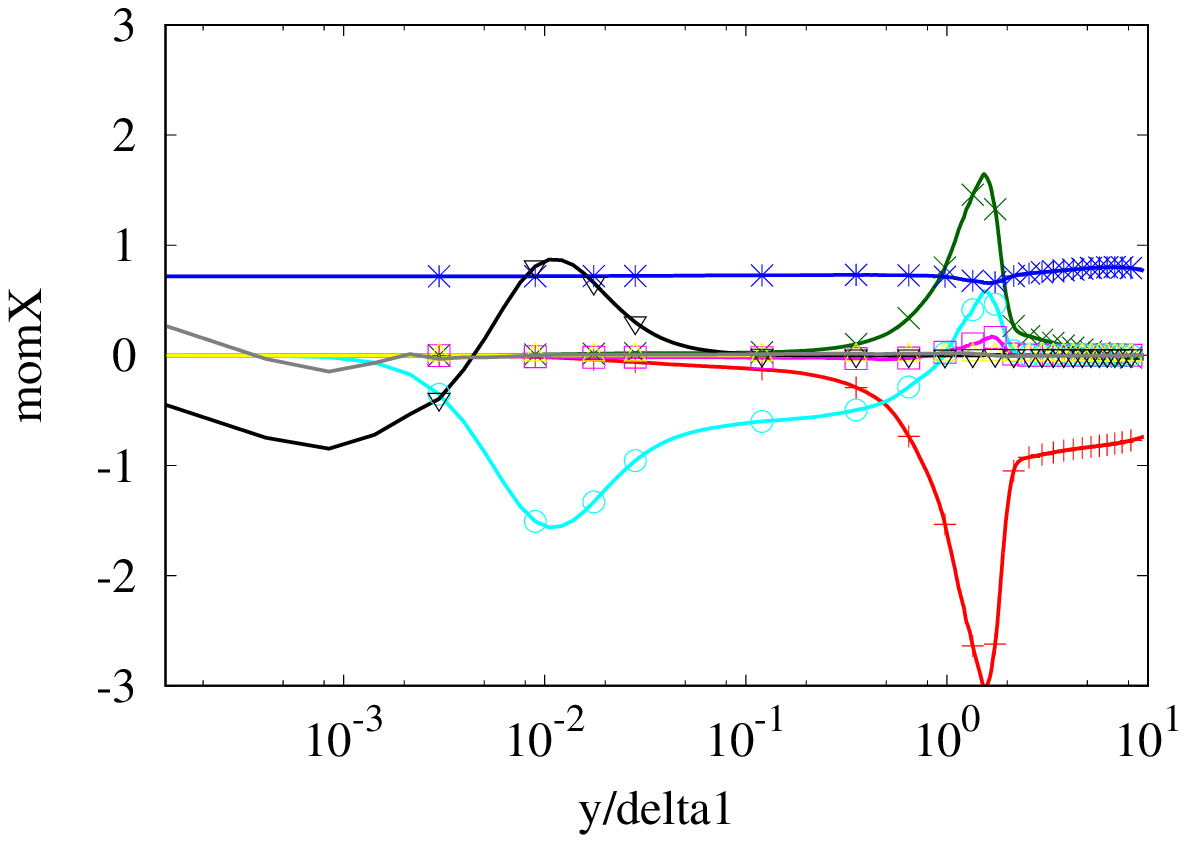} \\
  & \\
  (c) & (d) \\
  \psfrag{momY}[c][b][1.0]{$y$-momentum terms $\times10^2$}
  \includegraphics[trim = 11mm 0mm 6mm 0mm, clip, height=0.41\textwidth]{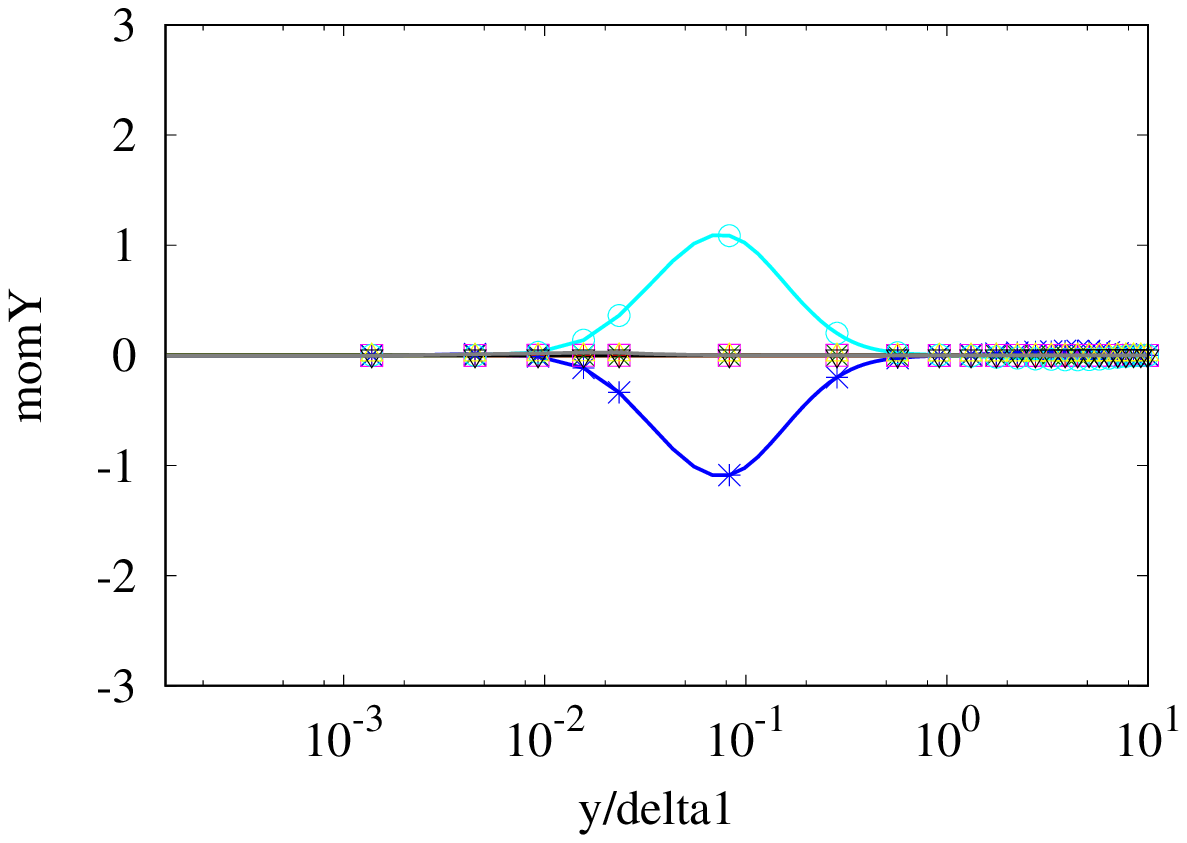} &
  \psfrag{momY}[c][b][1.0]{}
  \includegraphics[trim = 18mm 0mm 6mm 0mm, clip, height=0.41\textwidth]{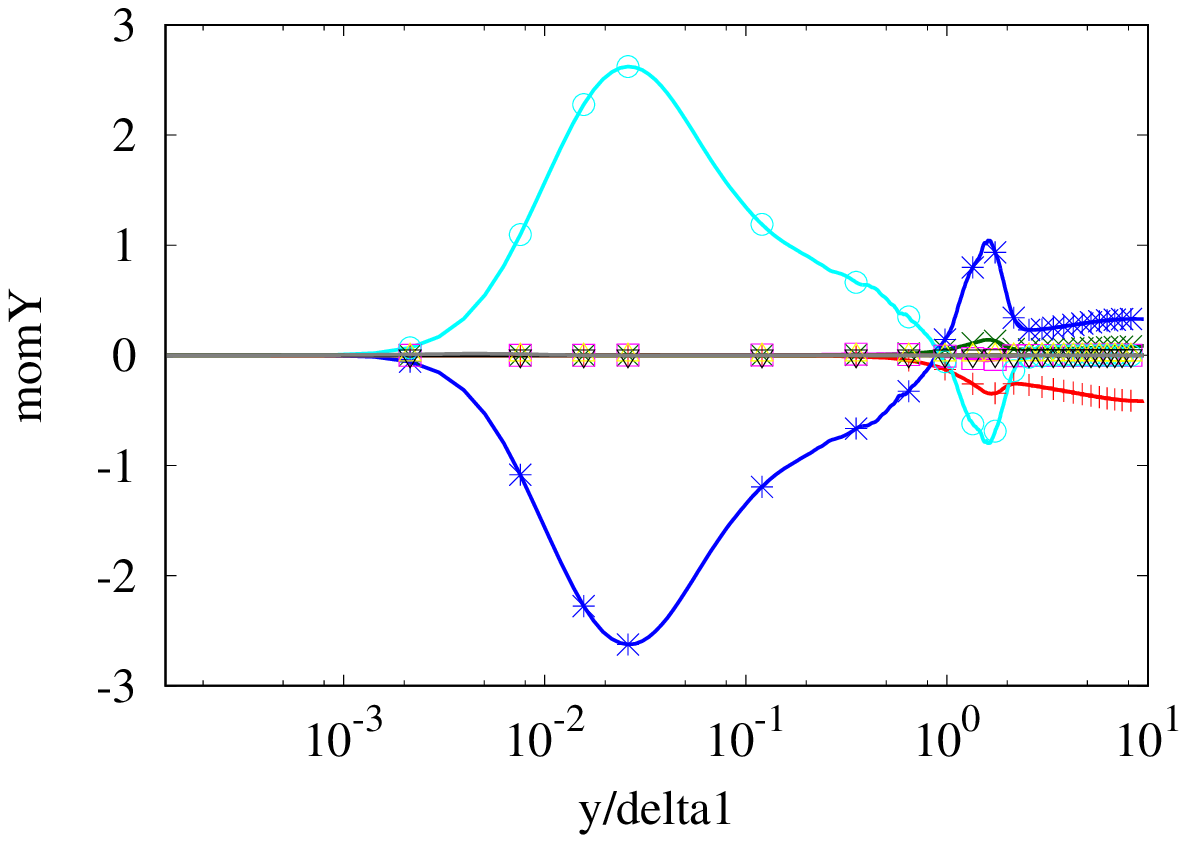} \\
  \end{tabular}
\end{center}
\caption{Streamwise averaged momentum term profiles nondimensionalised by $\delta_1$ and $U_e$.
The streamwise momentum equation terms
$\langle U \rangle \p_x \langle U \rangle$ (\textcolor{red}{$+$}), 
$\langle V \rangle \p_y \langle U \rangle$ (\textcolor{green}{$\times$}), 
$\p_x \langle P \rangle$ (\textcolor{blue}{$*$}), 
$\p_x \langle uu \rangle$ (\textcolor{magenta}{\opensquare}),
$\p_y \langle uv \rangle$ (\textcolor{cyan}{$\circ$}), 
$-\nu \p_x \p_x \langle U \rangle$ (\textcolor{yellow}{\opentriangle}), 
$-\nu \p_y \p_y \langle U \rangle$ (\opentriangledown), and
the residual given by the negative sum of the aforementioned terms (grey) for the
(a) ZPG TBL; and
(b) strong APG TBL, with same vertical axis as (a).
The wall normal momentum equation terms
$\langle U \rangle \p_x \langle V \rangle$ (\textcolor{red}{$+$}),
$\langle V \rangle \p_y \langle V \rangle$ (\textcolor{green}{$\times$}),
$\p_y \langle P \rangle$ (\textcolor{blue}{$*$}),
$\p_x \langle uv \rangle$ (\textcolor{magenta}{\opensquare}),
$\p_y \langle vv \rangle$ (\textcolor{cyan}{$\circ$}),
$-\nu \p_x \p_x \langle V \rangle$ (\textcolor{yellow}{\opentriangle}),
$-\nu \p_y \p_y \langle V \rangle$ (\opentriangledown), and
the residual given by the negative sum of the aforementioned terms (grey) for the
(c) ZPG TBL; and
(d) strong APG TBL, with same vertical axis as (c).
}
\label{fig:mom}
\end{figure}

The relative magnitude (and arguably importance) of the terms in the Navier-Stokes equations is now assessed.
The terms in the streamwise RANS equation (\ref{eq:x-mom}) and the wall normal RANS equation (\ref{eq:y-mom}) are time and spanwise averaged, for the ZPG and the strong APG TBL.
The Reynolds stress gradients in these equations quantify the transfer of momentum between the fluctuating and mean fields.
The statistics in figure~\ref{fig:mom} are presented such that negative values of the Reynolds stress gradients represent a transfer of momentum from the mean field to the fluctuating field, with positive values representing the reverse transfer.
All terms are nondimensionalised on the basis of $\delta_1$ and $U_e$.
The limits on the independent and dependent axes are kept constant to facilitate a direct comparison between the statistics.

Firstly we compare the two TBLs on the basis of the relative magnitudes of the $x$ momentum terms.
For the ZPG TBL, figure~\ref{fig:mom}(a) illustrates that the positive viscous term ($-\nu \p_y \p_y \langle U \rangle$) is in balance with the negative Reynolds stress gradient ($\p_y \langle uv \rangle$), each with a single inner peak.
Here there is a net transfer of streamwise momentum from the mean field to the fluctuating field via the dominant Reynolds stress gradient.
The remaining $y$ momentum terms are negligible in comparison.
The $x$ momentum terms for the strong APG TBL are illustrated in figure~\ref{fig:mom}(b).
The obvious difference is the nonzero positive pressure gradient ($\p_x \langle P \rangle$) which is relatively constant in $y$.
In the inner region $-\nu \p_y \p_y \langle U \rangle$ is positive, $\p_y \langle uv \rangle$ is negative, and they are in balance with the pressure gradient.
Both $-\nu \p_y \p_y \langle U \rangle$ and $\p_y \langle uv \rangle$ are reduced in magnitude in comparison to their respective ZPG counterparts.
In the outer region the viscous term becomes negligible.
The $\p_y \langle uv \rangle$ term has a positive outer peak, with the sign changing from negative to positive at $y=\delta_1$.
This is consistent with $-\langle uv \rangle$ having a maximum at $y=\delta_1$, as observed in figure~\ref{fig:BL_stresses}(d).
Note that the momentum transfer in the inner region is the same as in the ZPG TBL.
However, in the outer region there is a net transfer of momentum from the fluctuating to the mean field.
In contrast to the ZPG statistics, the convective terms are no longer negligible, and are in fact dominant. 
This is due to the non-zero streamwise and wall normal derivatives of the velocity field, and a non-negligible $\langle V \rangle$ as a result of the decelerating velocity and zero vorticity boundary conditions detailed in section~\ref{sec:BCs}.
At the outer peak, the convective term $\langle U \rangle \p_x \langle U \rangle$ is negative and $\langle V \rangle \p_y \langle U \rangle$ is positive.

The two TBLs are now compared on the basis of the relative magnitudes of the $y$ momentum terms.
For the ZPG TBL, the negative $\p_y \langle P \rangle$ and positive $\p_y \langle vv \rangle$ are dominant and in balance, each with a single inner peak, as illustrated in figure~\ref{fig:mom}(c).
For the strong APG TBL illustrated in figure~\ref{fig:mom}(d), at the inner peak the negative $\p_y \langle P \rangle$ and positive $\p_y \langle vv \rangle$ are again dominant and in balance, with the remaining terms negligible.
At the outer peak these terms have the opposite sign.
The Reynolds stress gradient ($\p_y \langle vv \rangle$) has a negative outer peak with the sign change occurring at $y=\delta_1$.
This is consistent with $\langle vv \rangle$ being a maximum at this location, as illustrated in figure~\ref{fig:BL_stresses}(b).
The convective term $\langle U \rangle \p_x \langle V \rangle$ is also significant in the farfield, which is consistent with the increased mean field shear.

In summary the convective terms transition from being negligible in the ZPG TBL to dominant in the strong APG TBL in the outer region.
The enhanced convective terms are consistent with an increase in mean shear, which using the linear stability arguments from section~\ref{sec:Reynolds_stresses}, is advantageous for the enhanced generation of turbulent fluctuations.
In the inner region there is a net transfer of streamwise momentum from the mean to the fluctuating field, and a net transfer of wall normal momentum from the fluctuating to the mean field.
In the outer region the transfers are reversed.

\section{Kinetic energy budgets}
\label{sec:budget_profiles}

The generation, dissipation and transfer of turbulent fluctuations in each of the TBL are now further quantified by the kinetic energy budgets.
For flows in statistical steady state (i.e. time derivatives are zero) the kinetic energy budget is given by
\begin{eqnarray}
  0 &=& \mathcal{M} + \mathcal{Z} + \mathcal{T} + \mathcal{P} + \mathcal{D} + \mathcal{V} \ \mbox{, }
\end{eqnarray}
\noindent where $\mathcal{M}$ is the mean convection, $\mathcal{Z}$ pressure transport, $\mathcal{T}$ turbulent transport, $\mathcal{P}$
production, $\mathcal{D}$ is the pseudo-dissipation, and $\mathcal{V}$ the viscous diffusion.
Each term is defined as
\begin{eqnarray}
\label{eq:mean_convection}
  \mathcal{M} &=& - \langle U_j \rangle \p_{x_j} E \ \mbox{, } \\
  \mathcal{Z} &=& - \p_{x_i} \langle p u_i \rangle \ \mbox{, } \\
  \mathcal{T} &=& - \p_{x_j} \langle u_i u_i u_j \rangle / 2 \mbox{, } \\
  \mathcal{P} &=& - \langle u_i u_j \rangle \ \p_{x_j} \langle U_i \rangle \ \mbox{, } \\
  \mathcal{D} &=& - \nu \left\langle (\p_{x_j} u_i) \ (\p_{x_j} u_i ) \right\rangle \ \mbox{, } \\
\label{eq:viscous_diffusion}
  \mathcal{V} &=& = \nu \p_{x_j} \p_{x_j} E \ \mbox{, } 
\end{eqnarray}
\noindent where $E = \langle u_k u_k \rangle / 2$ is the kinetic energy.

\begin{figure}
  \psfrag{ZPG}[c][][1.0]{$\beta=0$}
  \psfrag{beta1}[c][][1.0]{$\beta=1$}
  \psfrag{betaInf}[c][][1.0]{$\beta=39$}
\begin{center}
  \begin{tabular}{ll}
  (a) & (b) \\
  \psfrag{R}[c][][1.0]{$\mathcal{R}$}
  \psfrag{V}[c][][1.0]{$\mathcal{V}$}
  \psfrag{M}[c][][1.0]{$\mathcal{M}$}
  \psfrag{T}[c][][1.0]{$\mathcal{T}$}
  \psfrag{Z}[c][][1.0]{$\mathcal{Z}$}
  \psfrag{D}[c][][1.0]{$\mathcal{D}$}
  \psfrag{P}[c][][1.0]{$\mathcal{P}$}
  \psfrag{budgets*1e3}[c][b][1.0]{APG $\beta=39$ budgets$\times10^3$}
  \psfrag{y/delta1}[c][b][1.0]{$y/\delta_1$}
  \includegraphics[trim = 10mm 0mm 3mm 0mm, clip, width=0.51\textwidth]{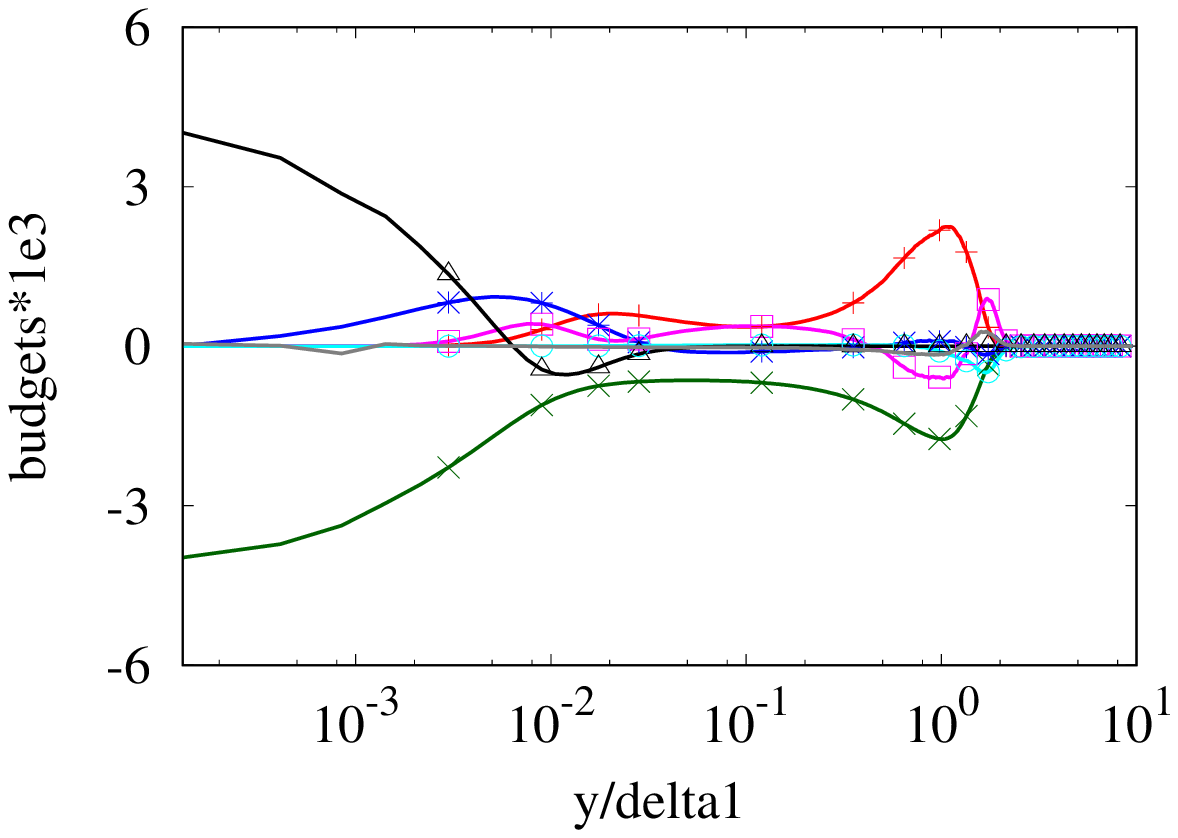} &
  \psfrag{budgets*1e3}[c][b][1.0]{production$\times10^3$}
  \psfrag{y/delta1}[c][b][1.0]{$y/\delta_1$}
  \includegraphics[trim = 10mm 0mm 3mm 0mm, clip, width=0.51\textwidth]{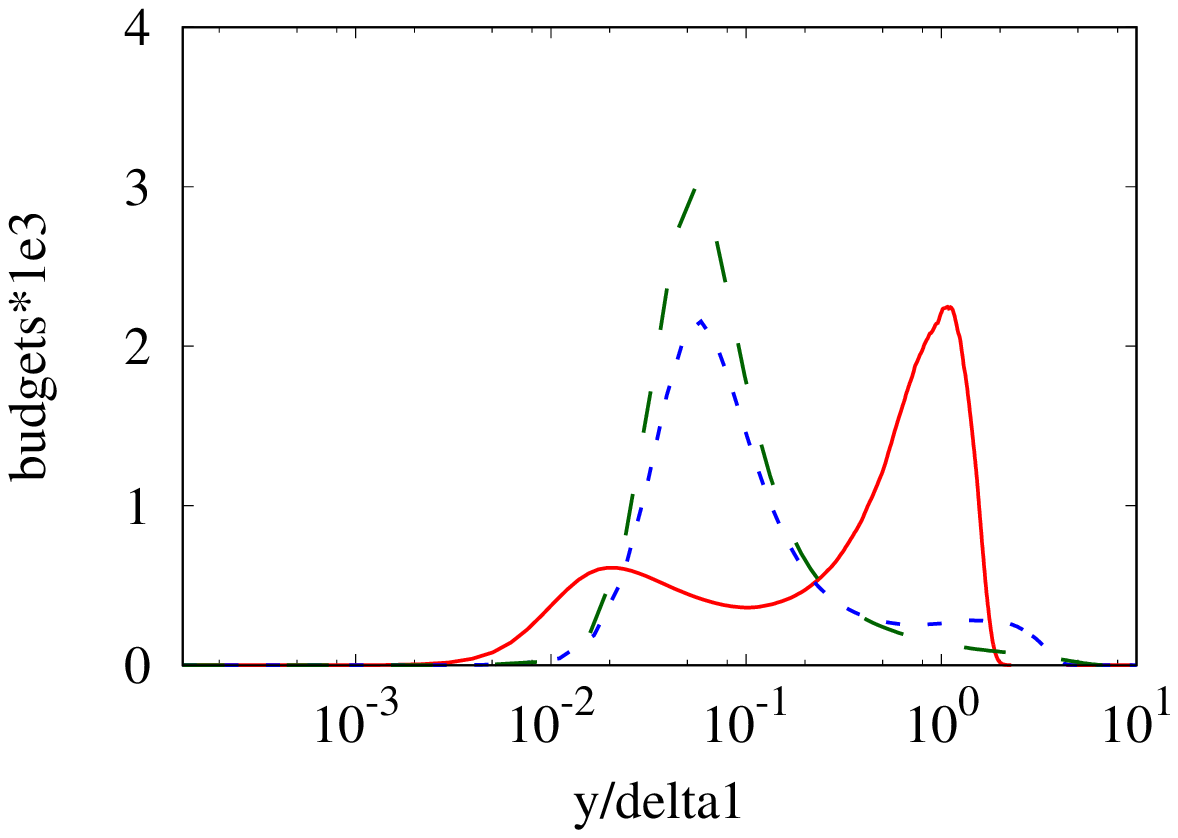} \\
  \end{tabular}
\end{center}
\caption{Streamwise averaged kinetic energy budget profiles nondimensionalised by $U_e$ and $\delta_1$.
(a) For the strong APG TBL, profiles of 
mean convection ($\mathcal{M}$, \textcolor{cyan}{$\circ$}),
pressure transport ($\mathcal{Z}$, \textcolor{blue}{$*$}),
turbulent transport ($\mathcal{T}$, \textcolor{magenta}{\opensquare}),
production ($\mathcal{P}$, \textcolor{red}{$+$}),
pseudo-dissipation ($\mathcal{D}$, \textcolor{green}{$\times$}), and
viscous diffusion ($\mathcal{V}$, \opentriangle), 
are all defined in equations~(\ref{eq:mean_convection}) to (\ref{eq:viscous_diffusion}) respectively, 
with the residual given by the negative sum of the aforementioned terms (grey line).
(b) Production profiles of the strong APG (solid red lines), mild APG (short dashed blue lines) and ZPG (long dashed green lines) TBL.
}
\label{fig:budgets}
\end{figure}

The terms in the kinetic energy budget are time and spanwise averaged, and then scaled using $U_e$ and $\delta_1$ as the pertinent velocity and length scale respectively.
Within the domain of interest these profiles are then additionally streamwise averaged in the scaled coordinates.
These streamwise averaged profiles are presented in figure~\ref{fig:budgets}(a) for the strong APG TBL. 
There is a clear outer peak in the production and dissipation terms located at $y=\delta_1$.
This indicates that turbulent kinetic energy produced in the outer flow is also locally dissipated.
The turbulent transfer term ($\mathcal{T}$) also gives insight as to the source of the fluctuations.
Negative values of $\mathcal{T}$ indicate that on average energy is leaving that wall normal position to be redistributed elsewhere, whilst positive values of $\mathcal{T}$ indicate that energy is being directed toward that position.
The most negative peak in $\mathcal{T}$ is located at $y=\delta_1$, with the turbulent transfer positive both above and below this wall normal location.
This is consistent with the view that at the point of inflection ($y=\delta_1$) a shear flow instability produces fluctuations (peak in $\mathcal{P}$) that are locally dissipated (negative peak in $\mathcal{D}$), and transferred to regions both closer to and further away from the wall (negative peak in $\mathcal{T}$).

The production term provides further information on the relative importance of the sources of turbulent kinetic energy.
The streamwise averaged production profiles from each simulation are compared in figure~\ref{fig:budgets}(b).
The ZPG TBL has one inner peak, with the mild and strong APG cases having both an inner and outer peak.
As mentioned previously these peaks coincide with the points of inflection in the respective mean streamwise velocity profiles.
When scaled in outer variables, the magnitude of the inner production peak decreases as the pressure gradient increases from ZPG, to mild APG, and finally to the strong APG TBL.
The inner production peak of the ZPG and mild APG cases are located at very similar distances from the wall.
This is because any Reynolds number effects between the ZPG and mild APG cases are minimised as they are streamwise averaged over the same $Re_{\delta_1}$ range.
Note the maximum $Re_{\delta_1}$ in the domain of interest for the strong APG case is over five times that of the ZPG and mild APG cases.
The higher the Reynolds number the smaller the near wall structures with respect to $\delta_1$, hence the inner peak being located closer to the wall.
The magnitude of the outer production peak increases with pressure gradient.
For the ZPG TBL there is no outer production peak, whilst for the mild APG TBL the outer peak is approximately one eighth the magnitude of the inner production peak.
In the strong APG TBL, the outer peak is three times the magnitude of the inner peak.
This transition in dominance from the inner production peak to the outer peak, is again consistent with the view that the flow is becoming less like a ZPG TBL and more like a free shear layer as the pressure gradient increases.

\section{Streamwise velocity spectra}
\label{sec:spanwise_spectra}

The relative contribution of the spanwise scales to the turbulent fluctuations throughout the wall normal domain, in particular at the outer peak, is determined from the streamwise velocity spectra.
The spectra are presented as a function of spanwise wavelength ($\lambda_z$) and wall normal position ($y$).
We compare the ZPG case to the strong APG case to accentuate the difference in the scaling of the spectra.
Three streamwise positions are presented for the ZPG and strong APG cases throughout the respective computational domains.
For the ZPG case, the displacement thickness based Reynolds numbers ($Re_{\delta_1}$) are
$2.07\times10^3$ (black dotted),
$3.55\times10^3$ (magenta dashed), and 
$4.84\times10^3$ (cyan solid).
For the strong APG TBL, the associated $Re_{\delta_1}$ are
$1.10\times10^4$ (black dotted),
$1.84\times10^4$ (magenta dashed), and
$2.55\times10^4$ (cyan solid).
At each streamwise position the variance in the wall normal/spanwise wavelength plane is illustrated at three contour levels set to $0.3$, $0.5$ and $0.7$ of the maximum variance in that plane.

\begin{figure}
\centering
  \begin{tabular}{ll}
  (a) & (b) \\
  \includegraphics[trim = 8mm 0mm 40mm 10mm, clip, width=0.5\textwidth]{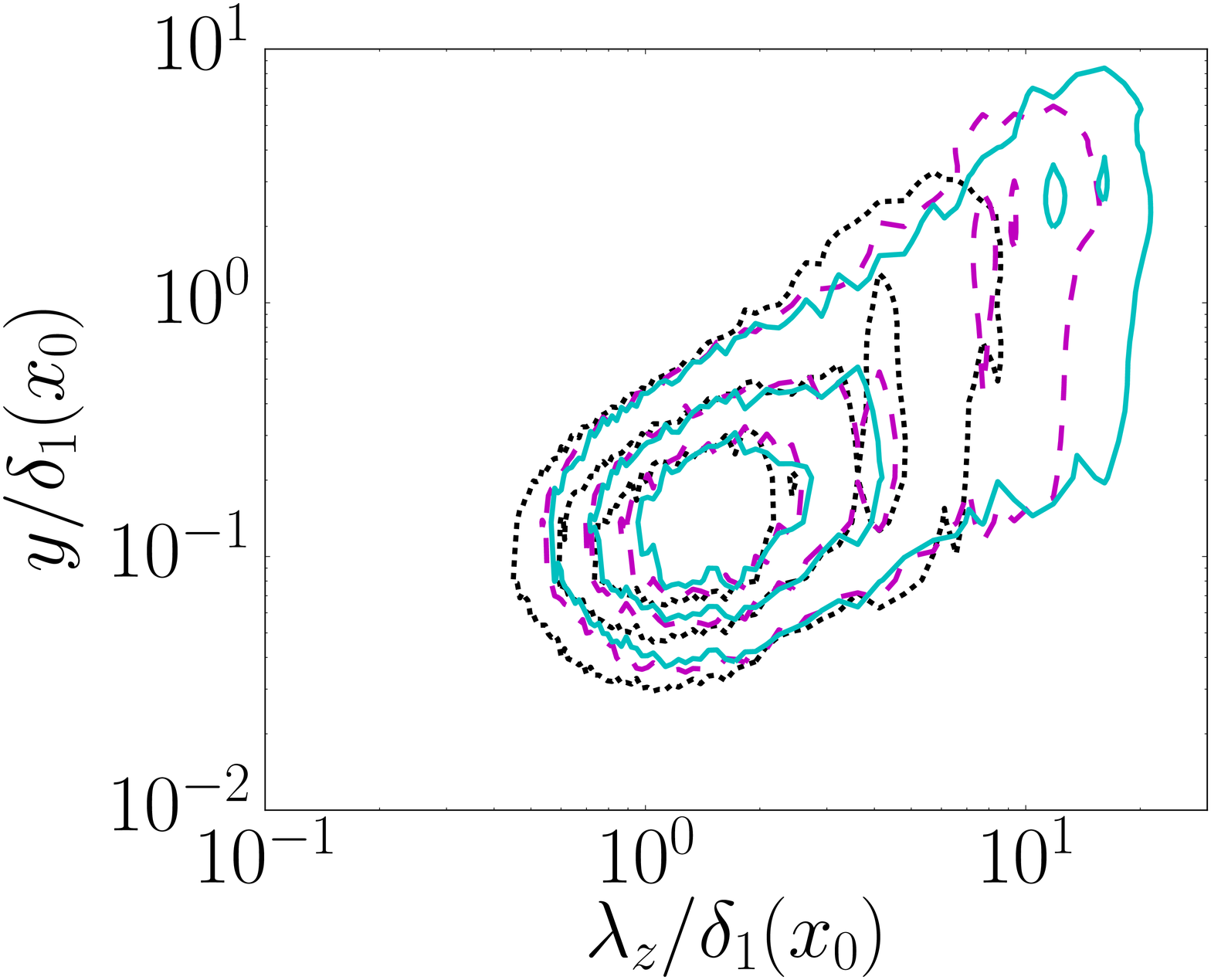} &
  \includegraphics[trim = 8mm 0mm 40mm 10mm, clip, width=0.5\textwidth]{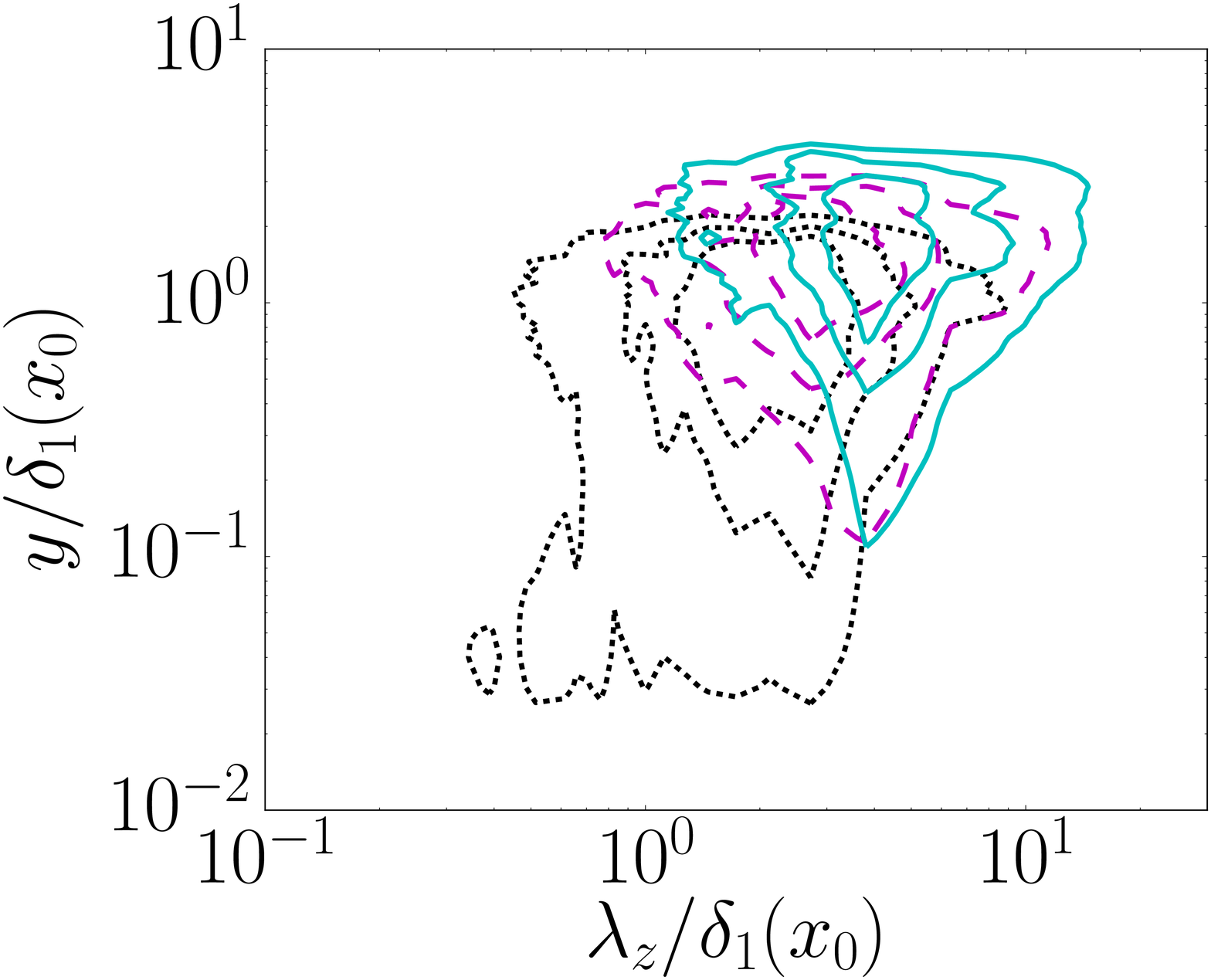} \\
  (c) & (d) \\
  \includegraphics[trim = 8mm 0mm 40mm 10mm, clip, width=0.5\textwidth]{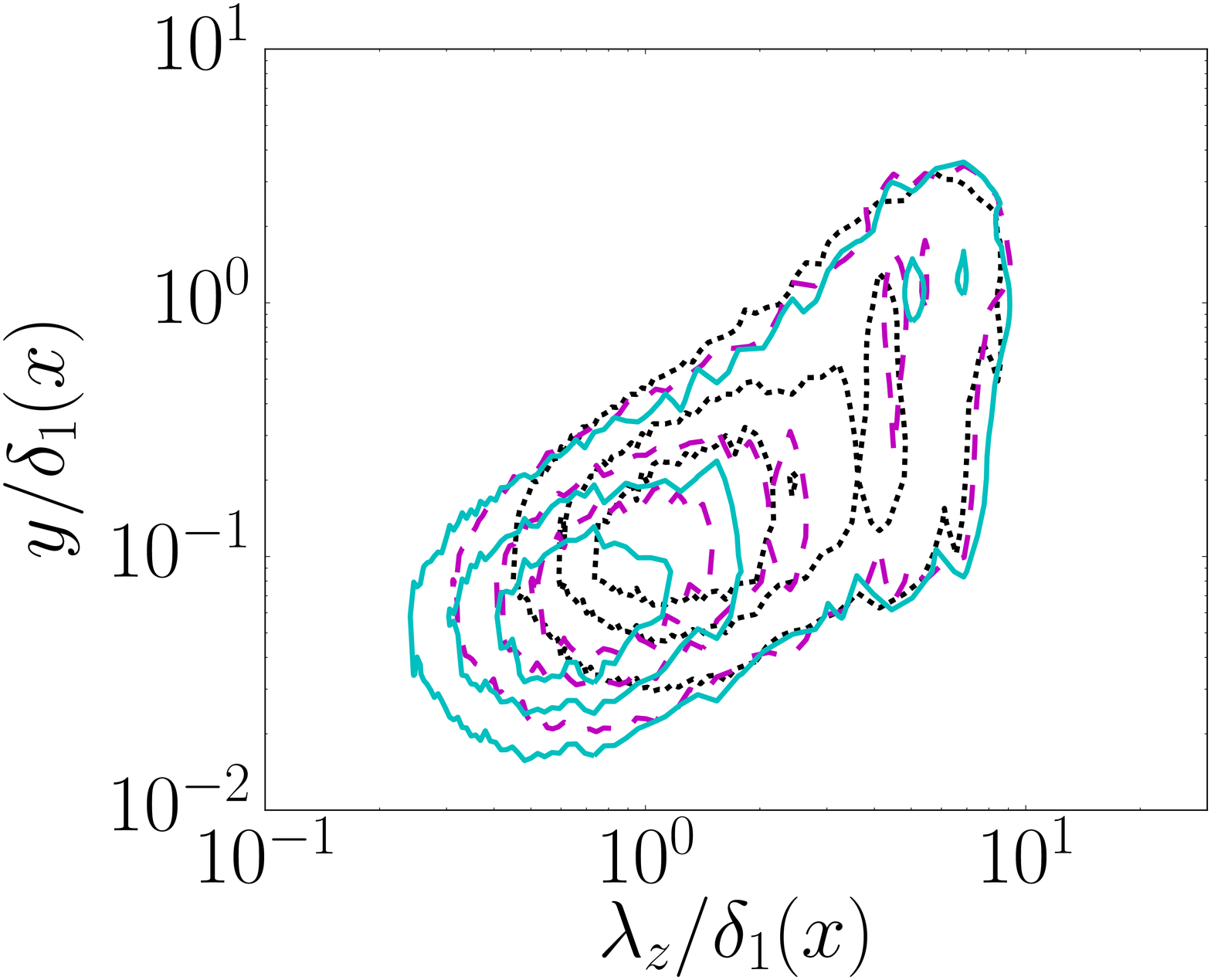} &
  \includegraphics[trim = 8mm 0mm 40mm 10mm, clip, width=0.5\textwidth]{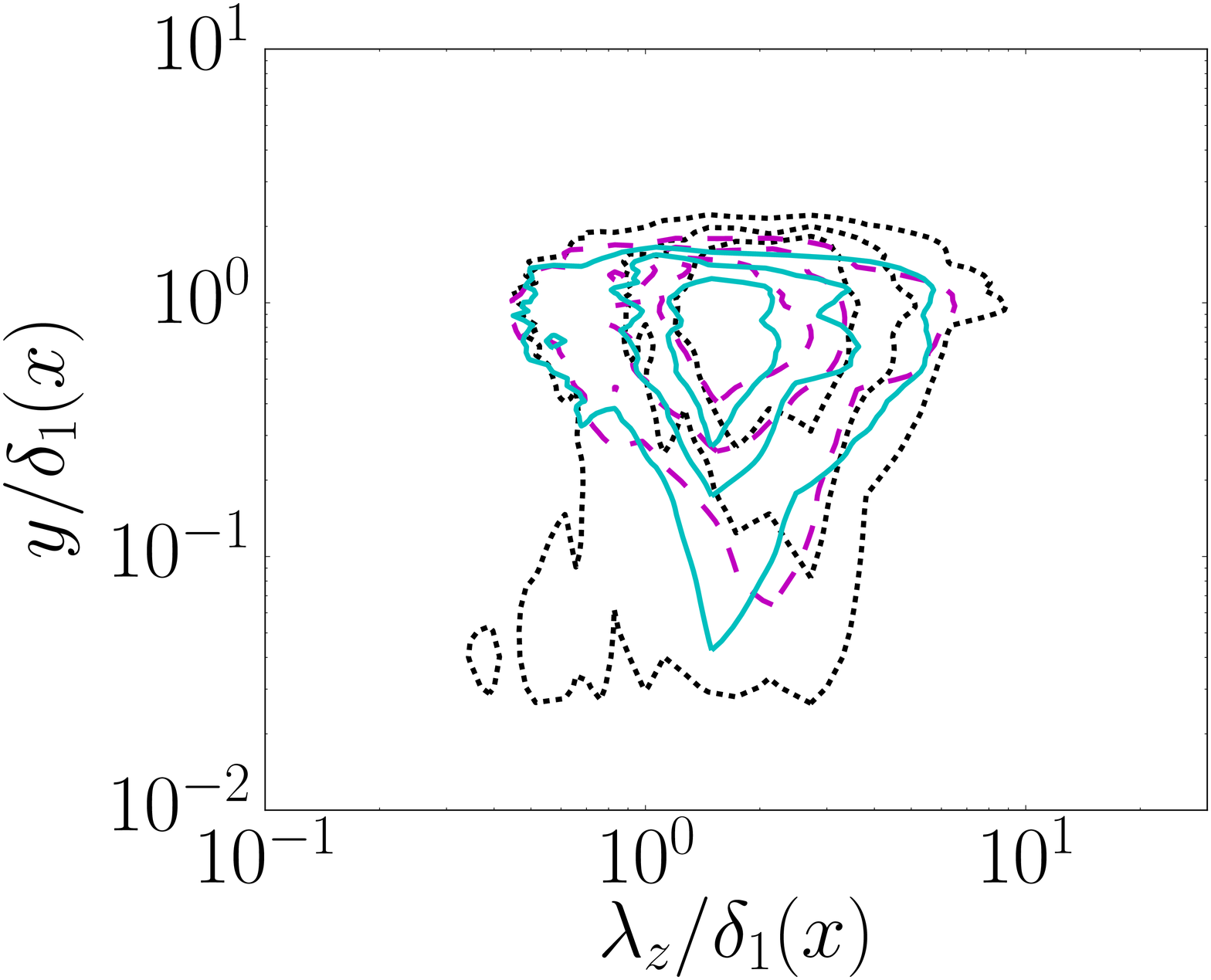} \\
  \end{tabular}
\caption{Streamwise velocity spectra at various streamwise positions for the ZPG and strong APG TBL.
In the wall normal/spanwise wavelength plane scaled by the displacement thickness at the position of the first spectrum:
(a) ZPG TBL; and
(b) strong APG TBL.
In the wall normal/spanwise wavelength plane scaled by the local displacement thickness for the:
(c) ZPG TBL; and
(d) strong APG TBL.
Contour levels are $0.3$, $0.5$ and $0.7$ times the maximum variance in the ($y$ , $\lambda_z$) plane for each streamwise position.
The ZPG TBL positions are
$Re_{\delta_1}=2.07\times10^3$ (black dotted),
$Re_{\delta_1}=3.55\times10^3$ (magenta dashed), and 
$Re_{\delta_1}=4.84\times10^3$ (cyan solid).
The strong APG TBL positions are
$Re_{\delta_1}=1.10\times10^4$ (black dotted),
$Re_{\delta_1}=1.84\times10^4$ (magenta dashed), and
$Re_{\delta_1}=2.55\times10^4$ (cyan solid).
}
\label{fig:spectra}
\end{figure}

In figure~\ref{fig:spectra}(a) the ZPG spectra are plotted against $y$ and $\lambda_z$, both scaled by the $\delta_1$ at the first streamwise station (of $Re_{\delta_1}=2.07\times10^3$).
The location of the maximum variance is positioned in the near wall region, and relatively independent of streamwise position.
As one moves downstream and the boundary layer thickens, the outer spectral peak moves further away from the wall and to larger spanwise wavelengths.
Note when plotted in viscous units the inner peak of the ZPG spectra collapse.
We chose not to present the data in this manner since the viscous lengths scale, $l^+ \equiv \nu/U_\tau$, becomes undefined for the incipient separation APG TBL.
As done above for the ZPG case, in figure~\ref{fig:spectra}(b) the spectra of strong APG TBL is plotted against $y$ and $\lambda_z$ scaled by the $\delta_1$ at the first streamwise station (of $Re_{\delta_1}=1.10\times10^4$).
Here the outer peak is dominant, with the location of the maximum variance strongly dependent upon the streamwise position.
When $y$ and $\lambda_z$ are scaled by the local $\delta_1$ the location of the outer peak collapses for different streamwise positions of the ZPG TBL in figure~\ref{fig:spectra}(c), and of the strong APG TBL in figure~\ref{fig:spectra}(d).
For the strong APG case the maximum variance is located at a wall normal position of $\delta_1$ and with a spanwise wavelength of approximately $2\delta_1$.
Note, a structure with a spanwise width of $\delta_1$, separated in the spanwise direction by a distance of $\delta_1$ to the next similar structure, would have a dominant spanwise wavelength of $2\delta_1$.

\section{Two-point correlations}
\label{sec:two-point_correlations}

To give an indication of the spatial coherence of the structures centred at the displacement thickness height, two-point correlations are calculated in the streamwise / wall normal plane for each of the TBL.
The two-point correlation function for the streamwise velocity component is defined as
\begin{eqnarray}
\label{eq:two-point_correlation}
  \rho_{uu}(x,y;\breve{x},\breve{y}) &=& \f{\langle u(x,y) \ u(\breve{x},\breve{y}) \rangle}{\sqrt{ \langle u^2(x,y) \rangle \ \langle u^2(\breve{x},\breve{y}) \rangle} } \ \mbox{,}
\end{eqnarray}
\noindent where $\breve{x}$ and $\breve{y}$ are the reference locations with respect to which the correlation is made, and the averaging is done over the spanwise direction and time.
There are analogous two-point correlation function definitions for the wall normal ($\rho_{vv}$) and spanwise ($\rho_{ww}$) velocity components.
In the analysis to follow the streamwise reference position $\breve{x}$ is located in the middle of the respective domains of interest, and the wall normal reference position is located at $\breve{y}=\delta_{1}(\breve{x})$.
We select this wall normal position as it is in the vicinity of the maximum fluctuations of the Reynolds stresses in the APG cases.

\begin{figure}
\begin{center}
  \begin{tabular}{l}
    (a) \\
    \includegraphics[trim = 35mm 90mm 10mm 110mm, clip, width=0.92\textwidth]{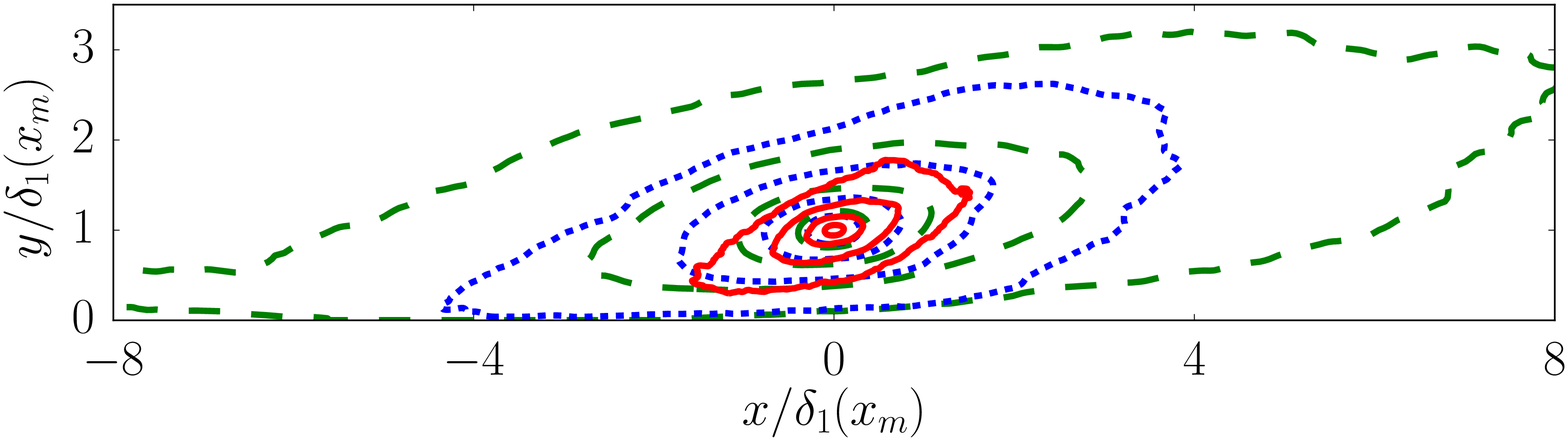} \\
    (b) \\
    \includegraphics[trim = 35mm 55mm 10mm 70mm, clip, width=0.92\textwidth]{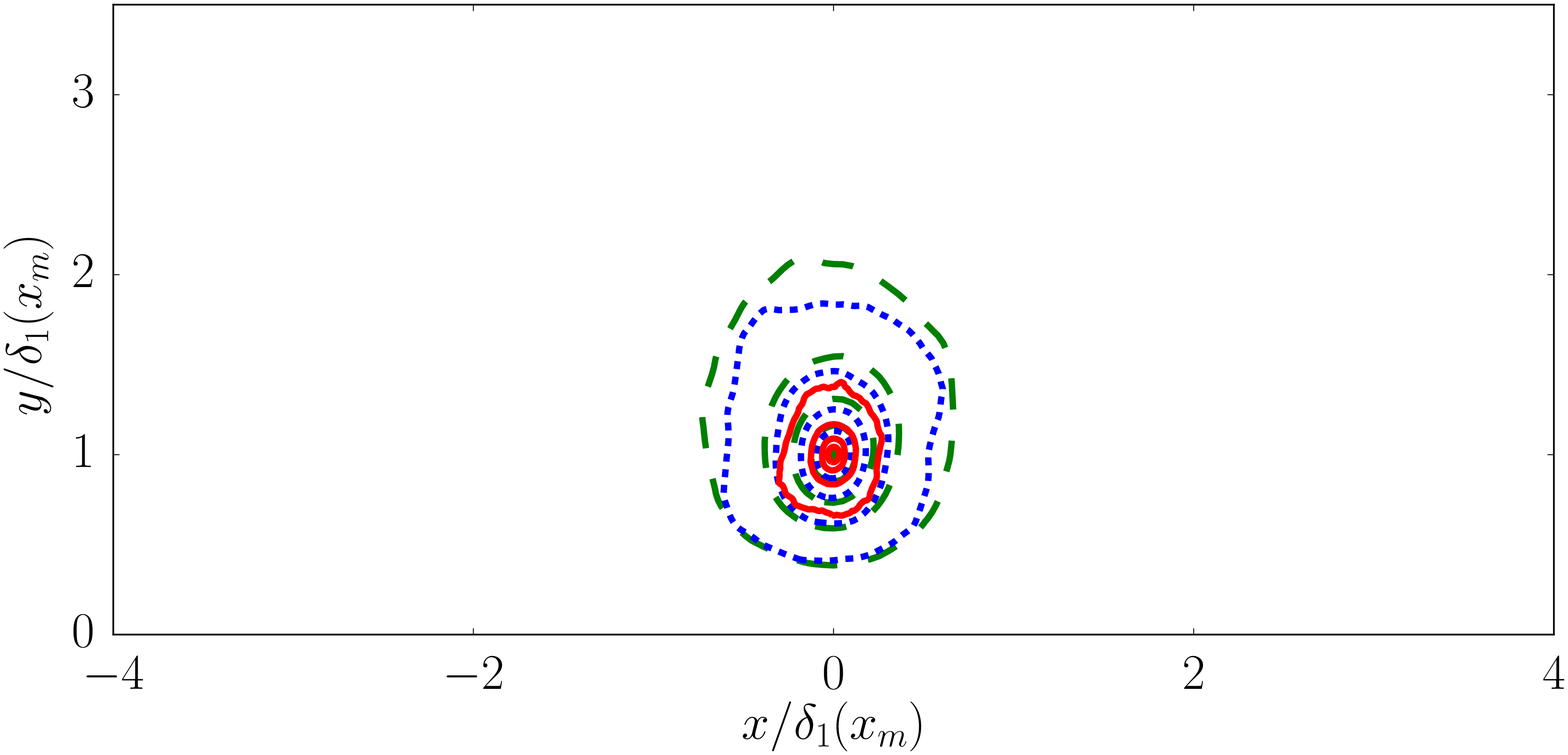} \\
    (c) \\
    \includegraphics[trim = 35mm 55mm 10mm 70mm, clip, width=0.92\textwidth]{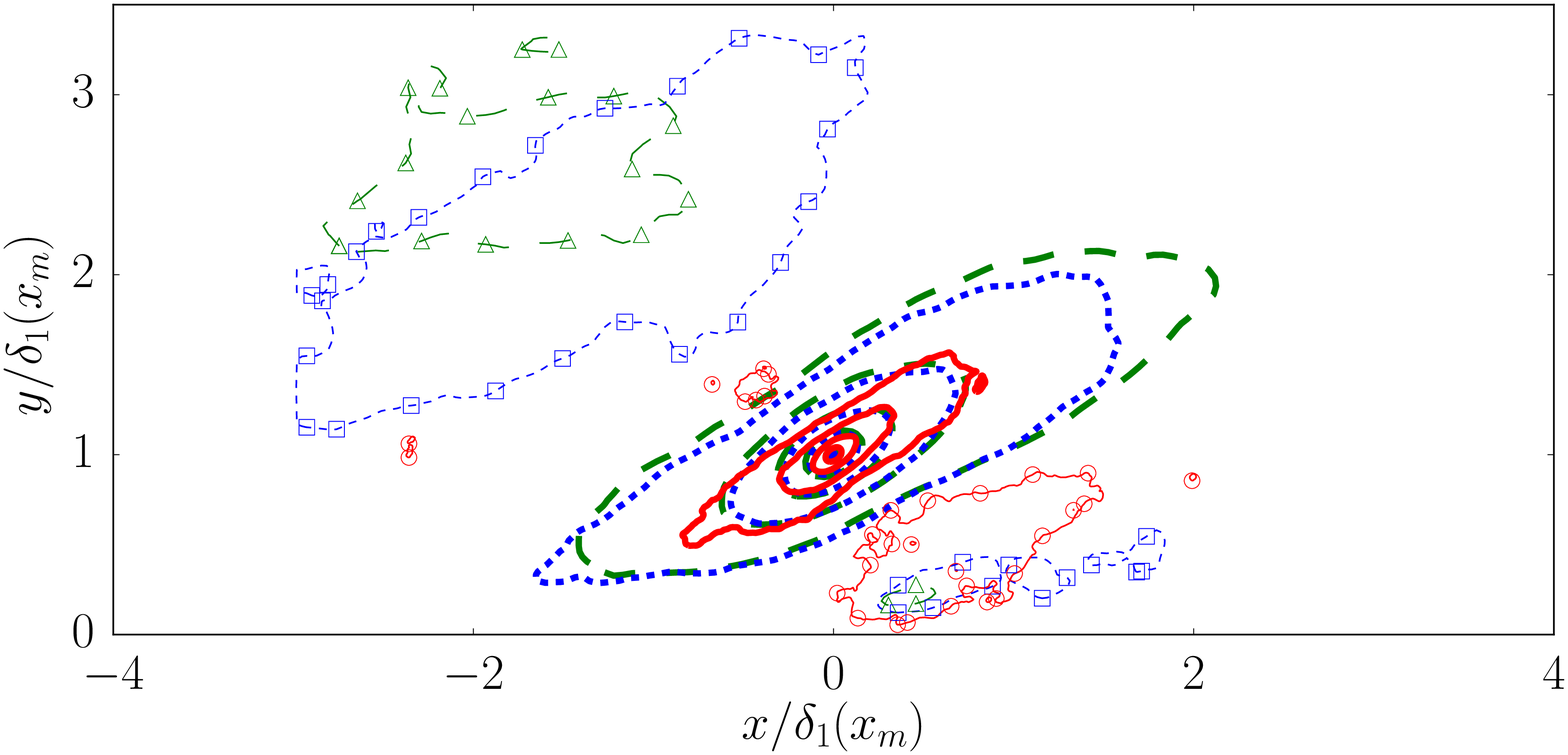} \\
  \end{tabular}
\end{center}
\caption{Two-point spatial correlation coefficients centred at $y=\delta_1$ and $x$ at the middle of the domain of interest for the
strong APG (solid red, positive correlation - thick lines, negative correlation - thin lines with \textcolor{red}{$\circ$}), 
mild APG (short dashed blue, positive correlation - thick lines, negative correlation - thin lines with \textcolor{blue}{\opensquare}) and 
ZPG (long dashed green, positive correlation - thick lines, negative correlation - thin lines with \textcolor{green}{\opentriangle}) TBL on the basis of:
(a) $\rho_{uu}$;
(b) $\rho_{vv}$; and
(c) $\rho_{ww}$.
In all figures the contour lines radiating out from the reference point are for values of $0.8$, $0.6$, $0.4$, $0.2$ and $-0.1$.
The horizontal and vertical axes are to scale.
The vertical axis has the same range in all figures.
Note the horizontal axis in (a) is twice the range of that in (b) and (c).
}
\label{fig:correlations}
\end{figure}

The correlation fields for $\rho_{uu}$, $\rho_{vv}$ and $\rho_{ww}$ are illustrated in figure~\ref{fig:correlations}(a), figure~\ref{fig:correlations}(b) and figure~\ref{fig:correlations}(c), respectively.
In each of these figures the horizontal and vertical axes are to scale in order to accurately visualise the aspect ratio and inclination of the correlation structures.
To facilitate a direct comparison between each of the correlation fields, the vertical axis has the same range in all figures, however, the horizontal axis in figure~\ref{fig:correlations}(a) is twice the range of that in figure~\ref{fig:correlations}(b) and figure~\ref{fig:correlations}(c).
The green long dashed contour lines represent the ZPG TBL, the blue short dashed contours the mild APG TBL, and the solid red contours the strong APG TBL.
The thick contour lines in order radiating out from the reference point are $0.8$, $0.6$, $0.4$, and $0.2$.
The thin contour lines with the symbols in figure~\ref{fig:correlations}(c) represent a contour value of $-0.1$.

For each TBL the $\rho_{uu}$ correlation field in figure~\ref{fig:correlations}(a) is elliptic in shape, with the ZPG contour lines extending further downstream and upstream from the reference point than the mild APG case, which in turn extends further back than the strong APG TBL.
The major axis of these elliptical structures is tilted upwards in the streamwise direction at an approximate angle of $7^\circ$ for the ZPG, $14^\circ$ for the mild APG, and $27^\circ$ for the strong APG TBL.
The $\rho_{uu}$ correlation field, therefore, becomes more compact and more inclined in the streamwise direction as the pressure gradient increases.
The correlation fields for $\rho_{vv}$ in figure~\ref{fig:correlations}(b) are not tilted in any particular direction, but also become more compact with increasing pressure gradient.
The $\rho_{ww}$ fields in figure~\ref{fig:correlations}(c) are elliptical in shape, slant upward in the streamwise direction, and are flanked by regions of negative correlation.
As the pressure gradient increases the $\rho_{ww}$ structures also become more compact.
The properties of the ZPG two-point correlations presented above are also consistent with those previously discussed in \cite{Sillero2014_PF} and \cite{Sillero2014_PhDthesis}.

The observation that the correlation structures are more compact with increasing pressure gradient is consistent with the form of the Reynolds stress profiles.
Recall that the two-point correlations in figure~\ref{fig:correlations} are based upon the fluctuating velocity fields with a wall normal reference position of $\breve{y}=\delta_1$.
The outer peak in the Reynolds stress profiles represents the variance of the fluctuating velocity fields localised in the vicinity of $y=\delta_1$.
One would then expect that the width of the outer peak in the Reynolds stress profiles is proportional to the size of the wall-normal extent of the correlation structures, and would hence decrease with pressure gradient.
This is in fact the case.
For example the half-width of the outer peak in the $\langle uu \rangle$ profiles illustrated in figure~\ref{fig:BL_stresses}(a) is $0.6\delta_1$ for the strong APG case, as compared to $2\delta_1$ for the mild APG case.
The half-width is defined here as the distance between the location of the outer peak and the position further away from the wall at which the variance drops to half of its peak value.

\section{Concluding remarks}
\label{sec:conclusion}

We compared three turbulent boundary layers generated using direct numerical simulation: a ZPG ($\beta=0$); a mild APG ($\beta=1$); and a strong APG ($\beta=39$).
The coefficients quantifying the extent of self-similarity of each of the terms in the boundary layer equations were assessed.
For all but the viscous term, $C_\nu$, the streamwise standard deviation of the self-similarity coefficients were found to be less than $2.5\%$ of the associated streamwise average.
The absolute magnitude of $C_\nu$, and its magnitude relative to the other self-similarity coefficients (e.g. $C_\nu/C_{uu}$) decreases with increasing pressure gradient, indicating that the viscous term is becoming a weaker constraint.
Within the domain of interest, the strong APG TBL mean velocity and Reynolds stress profiles are shown to collapse under outer scaling.

The manner in which the properties of the mean and fluctuating fields of the boundary layers change with increasing pressure gradient were documented.
For the mean streamwise velocity field, the extents of the log-layer and viscous sub-layer decrease, and the wake region expands with increasing pressure gradient.
The zone of influence of the viscous length and velocity scales is therefore reduced. 
This is consistent with only the outer length and velocity scales being required to collapse the mean velocity and Reynolds stress profiles.
The Reynolds stresses of the APG TBL cases were shown to exhibit a second outer peak, coinciding with the outer point of inflection in the mean streamwise velocity profile, and is suggestive of a shear flow instability.
The outer peak becomes more pronounced and more spatially localised as the pressure gradient increases.
Consistent with this increased localisation of the Reynolds stresses, two-point correlations of the velocity field centred at this outer peak illustrate that the statistical structures become more compact.
For the strong APG TBL, the streamwise velocity spectra were shown to also collapse in outer scaling, with the outer peak located at $y=\delta_1$ of dominant spanwise wavelength $\lambda_z=2\delta_1$.
At this outer peak there is a net transfer of streamwise momentum from the fluctuating to the mean field, and a transfer of wall normal momentum from the mean to the fluctuating field.
The momentum transfers are reversed for the inner peak.
The turbulent production term of all TBL exhibit inner peaks associated with the near wall shear, and the APG TBL also exhibit a second outer peak associated with the shear imparted as a result of the pressure gradient.
The outer production peak is non-existent in the ZPG TBL, small relative to the inner peak in the mild APG TBL, and dominant in the strong APG TBL.
The turbulent transfer term for the strong APG TBL has a negative peak at $y=\delta_1$ surrounded by positive turbulent transfer both above and below.

The above observations have lead to the following physical model of APG TBL.
The application of an APG imparts additional farfield shear resulting in a point of inflection in the mean streamwise velocity profile, given the APG is sufficiently strong.
A shear flow instability at this point of inflection locally generates turbulent kinetic energy which is transferred to regions both closer to and further away from the wall.
Likewise instabilities are also generated as a result of the wall shear.
The combined instabilities generate Reynolds stresses, and the mean field is then modified by a momentum transfer via the Reynolds stress gradients.
This modified mean field will generate a modified set of instabilities, producing modified Reynolds stress gradients, and so the cycle continues.
Finally, the above observations of the meanfield, Reynolds stresses, and production profiles, all indicate that as the pressure gradient increases the flow becomes less like a ZPG TBL (no mean farfield shear) and more like a free shear layer (no mean wall shear).

\section*{Acknowledgements}

The authors would like to acknowledge the research funding from the Australian Research Council and the European Research Council, and the computational resources provided by the Australian National Computational Infrastructure, iVEC and PRACE.
Javier Jimenez acknowledges the European Research Council grant ERC-2014.AdG-669505.
Julio Soria gratefully acknowledges the support of an Australian Research Council Discovery Outstanding Researcher Award fellowship.

\bibliographystyle{jfm}
\bibliography{0references}

\end{document}